\documentclass[useAMS, usenatbib, usegraphicx]{mn2e}

\title[Formation of Isolated Ellipticals]
{Formation, Evolution and Properties of Isolated Field Elliptical Galaxies}

\author[Sami-Matias Niemi et al.]
{
Sami-Matias Niemi$^{1,2}$\thanks{E-mail: niemi@stsci.edu (SMN)}, 
Pekka Hein\"am\"aki$^{2}$,
Pasi Nurmi$^{2}$ and
Enn Saar$^{3}$
\\
$^{1}$STScI, 3700 San Martin Drive, Baltimore, MD 21218, United States\\
$^{2}$University of Turku, Department of Physics and Astronomy, Tuorla
Observatory, V\"ais\"al\"antie 20, Piikki\"o, Finland\\
$^{3}$Tartu Observatory, EE-61602 T\~oravere, Estonia\\
}

\begin{document}

\date{Released}

\pagerange{\pageref{firstpage}--\pageref{lastpage}} \pubyear{200x}

\maketitle

\label{firstpage}

\begin{abstract}
We study the properties, evolution and formation mechanisms of isolated field
elliptical galaxies. We create a 'mock' catalogue of isolated field elliptical
galaxies from the Millennium Simulation Galaxy Catalogue, and trace their merging
histories. The formation, identity and assembly redshifts of simulated isolated
and non-isolated elliptical galaxies are studied and compared. Observational and
numerical data are used to compare age, mass, and the colour-magnitude relation.
Our results, based on simulation data, show that almost seven per cent of all
elliptical galaxies brighter than $-19$ mag in $B-$band can be classified as
isolated field elliptical galaxies. Results also show that isolated elliptical
galaxies have a rather flat luminosity function; a number density of $\sim 3
\times 10^{-6}h^{3}$ Mpc$^{-3}$ mag$^{-1}$, throughout their $B-$band magnitudes.
Isolated field elliptical galaxies show bluer colours than non-isolated
elliptical galaxies and they appear younger, in a statistical sense, according to
their mass weighted age. Isolated field elliptical galaxies also form and
assemble at lower redshifts compared to non-isolated elliptical galaxies. About
$46$ per cent of isolated field elliptical galaxies have undergone at least one
major merging event in their formation history, while the same fraction is only
$\sim 33$ per cent for non-isolated ellipticals. Almost all ($\sim 98$ per cent)
isolated elliptical galaxies show merging activity during their evolution,
pointing towards the importance of mergers in the formation of isolated field
elliptical galaxies. The mean time of the last major merging is at $z \sim 0.6$
or $6$ Gyrs ago for isolated ellipticals, while non-isolated ellipticals
experience their last major merging significantly earlier at $z \sim 1.1$ or $8$
Gyrs ago. After inspecting merger trees of simulated isolated field elliptical
galaxies, we conclude that three different, yet typical formation mechanisms can
be identified: solitude, coupling and cannibalism. Our results also predict a
previously unobserved population of blue, dim and light galaxies that fulfill
observational criteria to be classified as isolated field elliptical galaxies.
This separate population comprises $\sim 26$ per cent of all IfEs.
\end{abstract}


\section[]{Introduction}\label{Intro}

Galaxies reside in three different environments: clusters, groups and voids. The
majority, more than 50 per cent, of galaxies are found in groups and clusters
\citep{Humason:1956p593,Huchra:1982p1,Geller:1983p576,Nolthenius:1987p357,Ramella:2002p514}.
This is particularly true for elliptical galaxies, as the morphology-density
relation has shown \citep{Oemler:1974p646, Dressler:1980p645}. The merger
hypothesis \citep{Toomre:1972p619} suggests that the product of the merger of two
spiral galaxies will be an elliptical galaxy. If this holds, then the probability
to find an elliptical galaxy grows in environments with high densities and low
velocity dispersions, e.g. groups of galaxies. Therefore, local environment is
thought to play a crucial role in galaxy formation and evolution
\citep[e.g.][]{Farouki:1981p656, Balogh:1999p657, Kauffmann:1999p658,
Lemson:1999p654, Moore:1999p653}.

In observations it is relatively easy to study cluster and group galaxies.
Because of this and the potential formation mechanism of elliptical galaxies, it
is not surprising that most studied ellipticals are found in clusters and groups.
Therefore, the detailed properties of isolated field elliptical (IfE) galaxies
have not been extensively studied or very well understood; there are only a few
observational studies and the surveys are small. Moreover, the formation
mechanisms and evolutionary paths of these ``lonely'' elliptical galaxies are not
yet well understood.

In the past two decades several observational projects have identified and
studied the properties of isolated field elliptical galaxies \citep[e.g.][and
references therein]{Smith:2004p250, Reda:2004p502, HernandezToledo:2008p648,
Norberg:2008p652}. In many studies, based on observational data, different
possible formation scenarios have been proposed \citep[see
e.g.][]{Mulchaey:1999p563, Reda:2004p502, Reda:2007p559}, ranging from clumpy
collapse at an early epoch to multiple merging events. Also equal-mass mergers of
two massive galaxies or collapsed groups have been suggested. Theoretical studies
predict that isolated ellipticals are formed in relatively recent mergers of
spiral galaxy pairs, while large isolated ellipticals may be the result of
merging of a small group of galaxies \citep[e.g.][]{Jones:2000p624,
DOnghia:2005p128}.

Observational studies have shown that several IfEs reveal a number of features
such as tidal tails, dust, shells, discy and boxy isophotes and rapidly rotating
discs \citep[e.g.][]{Reduzzi:1996p622, Reda:2004p502, Reda:2005p561,
Hau:2006p557, HernandezToledo:2008p648} indicating recent merger and/or accretion
events. \citet{HernandezToledo:2008p648} concluded that at least $78$ per cent of
their isolated elliptical galaxies show some kind of mophological distortion, and
suggested that these galaxies suffered late dry mergers.
\citet{Reda:2004p502, Reda:2005p561} compiled a sample of $36$ candidates of
isolated early-type galaxies and studied their properties, and concluded that a
major merger of two massive galaxies could explain most observed features. They
also concluded that a collapsed poor group of a few galaxies is a possible
formation scenario. However, \citet{Marcum:2004p572} studied a similar sample of
isolated early-type galaxies, and concluded that isolated systems are
underluminous by at least a magnitude compared with objects identified as merged
group remnants. \citet{Reda:2007p559} concluded that mergers at different
redshifts of progenitors of different mass ratios and gas fractions are needed to
reproduce the observed properties of IfEs.

However, some IfEs have not shown any evidence of recent merging activity
\citep[e.g.][]{Aars:2001p564, Denicolo:2005p568}. \cite{Aars:2001p564}, who
studied a sample of nine isolated elliptical galaxies, identified five galaxies
that were located in environments similar to those of loose groups, while the
environments of the remaining four galaxies were confirmed to be on low density.
All galaxies showed smooth, azimuthally symmetric profiles, with no obvious
indications of dust lanes, nascent spiral structure or star-forming regions. It
is possible that the merging events have happened in distant past, and all signs
of merging events have been wiped out. \cite{Mihos:1995p596} found out, by using
a combination of numerical simulation and synthesized Hubble Space Telescope
(HST) Wide-Field Planetary Camera 2 (WFPC2) images, that merger remnants appear
morphologically indistinguishable from a ``typical'' elliptical $\leq 1$ Gyrs
after the galaxies merged, while \cite{Combes:1995p663} estimated from numerical
simulations that the time might be even less than $0.5$ Gyrs. Indeed, these times
are very short in time scales of galaxy evolution, making it difficult to find
observational evidence to back up formation via merging events. On the other
hand, it is equally possible that some or even all of these galaxies have
initially formed in underdense regions and developed quietly without any major
mergers.

Observational studies of isolated field elliptical galaxies use similar isolation
criteria as optical studies of fossil groups. Both classes show a large magnitude
gap between the first- and the second-ranked galaxy. However, there is no
criterion for IfEs that would require a presence of extended X-ray emission, as
in case of fossil groups. Additionally, fossil groups are not necessarily found
in low-density environments \citep{vonBendaBeckmann:2008p598} like IfEs. Despite
the differences, it is possible that the formation mechanisms of the two systems
are similar or related to each other. Therefore, it is interesting to compare
isolated field elliptical galaxies and fossil groups, and see if these systems
share a common origin.

In this paper, we use the Millennium Simulation \citep{Springel:2005p595}
together with a semi-analytical model \citep{DeLucia:2007p414} of galaxy
formation within dark matter haloes to identify isolated field elliptical
galaxies, to study their properties and formation history, and to compare them
with observational data. This paper is organised as follows. In Section
\ref{Sample}, we discuss the Millennium Simulation, the semi-analytical galaxy
catalogue used, and sample selection. In Section \ref{Results}, we compare the
properties of the Millennium IfEs with observations, while in Section
\ref{ResultsA} we show that the IfEs form a population that is different from the
regular ellipticals. In Section \ref{Formation} we concentrate on evolution of
IfEs and possible formation mechanisms, while in Section \ref{s:discussion} we
discuss our results, the formation of field elliptical galaxies and their
possible connection to fossil groups. Finally, in Section \ref{summary} we
summarise our results and draw the conclusions. Throughout this paper, we adopt a
parametrized Hubble constant: $H_{0} = 100 h$ km s$^{-1}$ Mpc$^{-1}$.

\section[]{Sample Selection}\label{Sample}


\subsection[]{The Millennium Simulation}\label{MS}

We use a simulation that covers a big enough spatial volume and has a
sufficient mass resolution -- the dark matter only Millennium Simulation
\citep[MS;][]{Springel:2005p595}, a $2160^{3}$-particle model of a co-moving cube
of size $500 h^{-1}$ Mpc, on top of which a publicly available semi-analytical
galaxy model \citep{DeLucia:2007p414} has been constructed. The cosmological
parameters used in the MS were: $\Omega_{m} = \Omega_{dm} + \Omega_{b} = 0.25$,
$\Omega_{b} = 0.045$, $h = 0.73$, $\Omega_{\Lambda} = 0.75$, $n = 1$, and
$\sigma_{8} = 0.9$ where the Hubble constant is characterized as $100 h$ km
s$^{-1}$ Mpc$^{-1}$ \citep[for detailed description of the MS,
see][]{Springel:2005p595}. These values were inferred from the first-year WMAP
(Wilkinson Microwave Anisotropy Probe) observations \citep{Spergel:2003p124}.


The galaxy and dark matter halo formation modeling of the MS data is based on
merger trees, built from $64$ individual snapshots. From these time-slices,
merger trees are built by combining the information of all dark matter haloes
found at any given output time. This enables us to trace the formation history
and growth of haloes and subhaloes through time. Properties of galaxies in the MS
data are obtained by using semi-analytic galaxy formation models (SAMs), where
star formation and its regulation by feedback processes is parametrized in
terms of analytical physical models. The semi-analytical galaxy catalogue we use
contains about nine million galaxies at $z = 0$ down to a limiting absolute
magnitude of $M_{R} - 5 \log h = -16.6$. A detailed description of the creation
of the MS Galaxy Catalogue, used in this study, can be found in
\citet{DeLucia:2007p414}.

Semi-analytical galaxy formation models are known to be less than perfect. There
are several free parameters in each SAM that all have a different impact on the
properties of galaxies. The free parameters usually control the feedback effects
and cooling of hot gas. The SAM also controls how gas is stripped from a galaxy
when it approaches another galaxy, and possible starburst events. The physics of
these processes described is not well known. However, in general, SAMs can
reproduce the observed galaxy luminosity function and other statistics well. As
our study is statistical by nature, it is likely that the SAM used for the
Millennium Simulation Galaxy Catalogue is accurate enough to predict the
properties of IfEs and their evolution and formation history.

The MS Galaxy Catalogue does not predict galaxy morphologies. To assign a
morphology for every galaxy, we use their bulge-to-disk ratios.
\citet{Simien:1986p431} found a correlation between the $B-$band bulge-to-disc
ratio and the Hubble type $T$ of galaxies from observations. The mean relation
may be written:
\begin{equation}\label{eq:T}
<\Delta m(T)> = 0.324x(T) - 0.054x(T)^{2} + 0.0047x(T)^{3} ,
\end{equation} 
where $\Delta m(T)$ is the difference between the bulge magnitude and the total
magnitude and $x(T) = T+5$. We classify galaxies with $T < -2.5$ as ellipticals,
those with $-2.5 < T < 0.92$ as S0s, and those with $T > 0.92$ as spirals and
irregulars. Galaxies without any bulge are classified as type $T = 9$.

\subsection[]{Selection of IfEs}\label{Selection}

There are a few different selection criteria that are being used to define
isolated field elliptical galaxies in observations \citep[see
e.g.][]{Colbert:2001p574, Marcum:2004p572, Reda:2004p502, Collobert:2006p580}.
Despite the differences in selection criteria, all studies identifying IfEs share
a common ideology. In general, IfEs are defined as elliptical galaxies that do
not have nearby optically bright companion galaxies, usually in the $B-$band.
This is tested by using an isolation sphere or a cone and choosing a minimum
magnitude difference between the brightest and the second brightest galaxy
$\Delta m_{12}$ inside the sphere or the cone. In observational studies, the
isolation cone is often taken as a circle in the sky that expands in redshift
space. Due to peculiar velocities, accurate line-of-sight distances of galaxies
are unknown complicating the identification. Thus, observational studies differ
in the numerical values of the isolation criteria. Despite these differences,
discussed next, we use all possible available observational data for comparison
as the number of observed targets is small. For completeness, we show in Section
\ref{S:Abundance} how the number of IfEs depends on the selection criteria.

\citet{Colbert:2001p574} adopted rather strict rules for isolation, as they
required that IfEs cannot have catalogued galaxies with known redshifts within a
projected radius of $1.0 h^{-1}$Mpc and a velocity of $\pm 1000$ km s$^{-1}$,
while for the morphology of the IfEs they required Hubble types $T \leq -3$.
However, as their source of galaxies was the Third Reference Catalogue of Bright
Galaxies \citep[RC3,][]{1991trcb.book.....D}, their sample IfEs might have faint
companion galaxies due to the incompleteness of the source catalogue. Missing
companions do not affect the study of bright isolated galaxies per se, but
their environment is affected by the incompleteness and missing companions.
\citet{Marcum:2004p572} based their initial sample of IfEs on the same catalogue.
They adopted an even stricter criterion in projected radius, a minimum projected
physical distance of $2.5$ Mpc to any nearest neighbour brighter than $M_{V} = -
16.5$. \citet{Reda:2004p502} used less strict criteria in their study and
required only that an IfE candidate has no neighbours with brightness difference
$\Delta m_{12} \leq 2.0$ mag, in the $B-$band, within $0.67$ Mpc in the plane of
sky and $700$ km s$^{-1}$ in recession velocity. Further, their elliptical
galaxies are also of Hubble type $T \leq -3$.

For the selection of IfEs from the Millennium Simulation we adopt the criteria
used by \citet{Smith:2004p250}, similar to those adopted by
\citet{Reda:2004p502}.
We concentrate on galaxies that are incontrovertibly ellipticals
and thus adopt a strict morphology limit $T \leq -4$. We further limit the brightness
of the IfE candidates with the magnitude cut-off of $M_B\leq -19$.
For the isolation criterion we adopt two isolation spheres with the radii of $0.5h^{-1}$
and $1.0 h^{-1}$Mpc. Within these isolation spheres, we require that the $B-$band
magnitude difference of the first- and second-ranked galaxies $\Delta m_{12}$
must be $\geq 2.2$ and $\geq 0.7$ mag, respectively. The application of the
criteria is illustrated in \citet[][Fig.$1$]{Smith:2004p250}.

Our adopted values of magnitude differences correspond to factors of about eight
and two in luminosity for the small and large sphere, respectively. This choice
ensures that possible companions are small and light enough not to produce any
major perturbations in the gravitational potential of the system. Note, that in
case of simulations our isolation criteria operate in real space rather than in
redshift space. Therefore, we can use isolation spheres rather than cones and we
are not plagued by interlopers due to inaccuracies in distance measurements.

\subsection[]{Simulated IfEs and Control Samples}

To identify isolated field elliptical candidates in the MS, we chose five
independent cubic volumes inside the simulation box. Initially, we chose each
volume to have a side length of $\sim 200h^{-1}$ Mpc, while none of the cubes
overlap. However, we later limited the volume from where the possible IfE
candidates can be identified to have a side length of $195h^{-1}$ Mpc. This was
done to enable the study of the surroundings of each IfE candidate in the same
fashion. Moreover, without the limitation, we could have accidentally identified
a field elliptical candidate $\leq 1.0 h^{-1}$Mpc from an edge of the box,
leading to a possible false identification. The five volumes were chosen to
overcome computational issues and for easier study of IfE environment. With five
independent volumes it is also possible to make comparisons between the IfEs of
each volume.

At first all galaxies inside each volume are treated as possible IfE candidates.
After identifying all candidates with $T \leq -4$ and $M_{B} \leq -19$ we apply
the isolation criteria discussed in the previous Section. This produces an
initial list of $302$ galaxies that fulfill the criteria for isolated field
elliptical galaxies. After compiling the initial list of IfEs, we inspected every
galaxy individually, and found that nine candidates were not the main galaxy of
the dark matter halo they reside in. Therefore, all galaxies that fulfill the
observationally motivated IfE criteria are not the central galaxies of their
Friends-of-Friends dark matter groups. In a strict sense these are not isolated
galaxies, as they belong to a dark matter halo containing galaxies more massive
and more luminous than the IfE candidate. Therefore, we omit them from the final
list of IfEs, which contains $293$ galaxies.

We also compile two control samples of elliptical galaxies for comparisons. The
first control sample contains in total $4563$ elliptical ($T \leq -4$) galaxies
brighter than $-19$ in the $B-$band at the redshift $z = 0$ and it is named as
Ellipticals (abridged as Es). For the second control sample, called Main
Ellipticals (abridged as MEs), we only select galaxies of the first control
sample that are the main galaxies of their dark matter haloes. This requirement
further limits the Ellipticals sample, and we are left with $1209$ galaxies in
the Main Ellipticals sample.


\subsubsection{Non-standard IfEs}

The removed galaxies are interesting from another point of view than regular
IfEs, as these nine non-standard IfEs are 'subhalo' galaxies (i.e. satellites of
a larger galaxy) and have multiple nearby companions, from $30$ to $80$ inside
the large isolation sphere. Thus, these galaxies are not isolated in galaxy
number density, but reside in a cluster rather than in the field. Physically it
is obvious that a satellite galaxy is not the dominant galaxy in its dark matter
halo. Since we find these non-standard IfEs, our observationally motivated
isolation criteria are not strict enough and the radii of the isolation spheres
should be slightly larger to avoid any false identifications. But as the number
of non-standard IfEs is small, we can keep the criteria and remove the
non-standard galaxies from our final sample, as described above.

The satellite non-standard ``isolated'' elliptical galaxies can shed some light
on the observational result \citep{Smith:2008p627} where an IfE galaxy (NGC 1600)
was also found to be located off from the dynamical centre of the system.
\cite{Sivakoff:2004p628} found that the X-ray emission is centered slightly to
the northeast of NGC 1600, suggesting that the galaxy is not at the centre of the
gravitational potential. The dynamical study and the X-ray observations together
suggest that NGC 1600 is surrounded by a massive halo extending out to several
hundred kiloparsecs. A subhalo galaxy, fulfilling the optical IfE criteria, could
explain the observations of NGC 1600. All nine galaxies in our simulations reside
inside a large dark matter halo, and in observations would be found to be off
from the centre of the potential well, as in the case of NGC 1600. However, a
massive and more luminous galaxy than NGC 1600 has not been detected at the
centre of the potential well in observations, while this is often the case in
simulations. Thus, a more straightforward explanation, where the IfE orbits
around the centre of the potential well and therefore seems to be shifted from
the centre, seen in X-ray observations, is more plausible. This is in agreement
with an X-ray study of another isolated elliptical NGC 4555, which is assumed not
to lie in the centre of a massive group-scale dark matter halo
\citep{OSullivan:2004p569}.

\section[]{Properties of observed and simulated field elliptical
galaxies}\label{Results}

Below we compare the properties of simulated IfEs with those of the observed
IfEs. If these are close enough, it will justify the theoretical study of the
formation, evolution and merger histories of IfEs, using the simulated IfE
population. Due to the large number of simulated IfEs we can actually make
theoretical predictions for properties, formation mechanisms and evolution of
IfEs that can be tested against observational data when the sample of
observations is large enough.

In following sections we use the Kolmogorov$-$Smirnov (KS) two-sample test to
study the maximum deviation between the cumulative distributions of two samples.
The null hypothesis of the KS test is that the two samples are from the same
population. We give our results as probabilities (p-values) that the difference
between the two samples could have arisen by change if they are drawn from the
same parent population.

\subsection[]{Number of IfEs}\label{S:Abundance}

The fraction of isolated field elliptical ($T \leq -4$ and $M_{B} \leq -19$ mag)
galaxies among all galaxies of any brightness, morphology or dark matter halo
status in the MS is merely $\sim 3.5 \times 10^{-3}$ per cent, corresponding to a
number density of $\sim 8.0 \times 10^{-6}$h$^{3}$ Mpc$^{-3}$. This is a very low
number density, however, such a straightforward calculation is not totally
justified. More meaningful number to conside is the fraction of IfEs among
elliptical galaxies. Moreover, our morphology limit ($T \leq -4$) for a simulated
IfE is very strict, biasing the fraction of elliptical galaxies to significantly
lower fractions than observed in the real Universe. Thus, we also identified
isolated field elliptical galaxies with relaxed morphology limit $T < -2.5$ and
quote the fractions below.

About $0.19$ per cent of all simulated elliptical (now $T < -2.5$) galaxies of
any brightness, can be identified as IfEs, when the criteria of Section
\ref{Selection} are adopted for IfEs. If we relax the strict morphology limit of
the IfEs and require that IfEs also have $T < -2.5$ the fraction rises to $\sim
0.48$ per cent as about $2.5$ times more IfEs can be identified. Relaxing the
morphology limit will therefore change the number density of IfEs to $\sim 1.9
\times 10^{-5}$h$^{3}$ Mpc$^{-3}$. When the brightness of the simulated
elliptical galaxies and their morphologies are limited to $M_{B} \leq -19$ at $z
= 0$ and $T \leq -4$, respectively, about $6.4$ per cent of these galaxies are
identified as IfEs at redshift zero. If we further limit our simulated elliptical
galaxies to galaxies that are the main galaxies of their dark matter haloes (as
in case of IfEs), $\sim 32$ per cent of ellipticals are now identified as IfEs.
Thus, IfEs are very rare objects among all galaxies, but at the same time, over
$30$ per cent of main elliptical galaxies brighter than $-19$ mag in $B-$band can
be classified as IfEs.

\cite{Stocke:2004p575} found from magnitude limited ($m_{B} \leq 15.7$)
observations that less than three per cent of all galaxies can be classified as
isolated. This value should be interpreted as an upper limit for the number of
IfEs, as it was based on all types of galaxies. \cite{Stocke:2004p575} further
found that early-type galaxies outnumber S$0$ galaxies $2:1$ in very isolated
areas. However, as spiral galaxies are the most dominant galaxy type in low
densities \citep[e.g.][]{Dressler:1980p645}, the observable estimate of the
number of IfEs is significantly lower than one per cent.
\citet{HernandezToledo:2008p648} concluded that early-type galaxies amount to
only $3.5$ per cent among all isolated galaxies, lowering the observational
estimate of IfEs to $\sim 1 \times 10^{-2}$ per cent. This fraction is in good
agreement with our findings as about $\sim 9 \times 10^{-3}$ per cent of all
simulated galaxies can be identified as IfEs when $T < -2.5$ morphology limit has
been adopted.

As observational studies adopt different criteria for isolation, we illustrate
their influence on the number of IfEs, with a stricter set of parameter values.
For simplicity, we adopt only one isolation sphere with a radius of $2.5h^{-1}$
Mpc, and require that the magnitude difference between the brightest and the
second brightest galaxy $\Delta m_{12}$ is $\geq 2.0$ mag inside the isolation
sphere. For the Hubble type, we adopt the same requirement as previously for our
IfEs: $T \leq -4$. Such values were used, e.g., by \citet{Marcum:2004p572}. With
these values we find that only $\sim 3 \times 10^{-4}$ per cent of all MS
galaxies can be classified as IfEs, corresponding to an extremely low number
density of $\sim 6 \times 10^{-7}$h$^{3}$ Mpc$^{-3}$. This is approximately ten
times less than before. It is obvious that changing the isolation criteria has a
major impact on the number of IfEs and therefore their statistical properties, as
stricter parameter values require a galaxy that is located in a true void. This
result also shows that galaxies with comparable luminosities tend to reside in
relatively close proximity rather than being truly isolated from any other
reasonable sized companion.

The results of this section show that the fraction of isolated field elliptical
galaxies depends highly on the isolation criteria adopted, but also on the
morphological type that is required. As the morphological type of simulated
galaxies is less than precise our number density for IfEs should not be taken as
a strict limit, but as a guideline when observations are being planned. Thus,
higher fractions of IfEs can be expected if isolation criteria or morphology
limits are being relaxed.

\subsection[]{Colour-magnitude Diagrams}\label{Comparison}

\begin{figure}
\includegraphics[width=84mm]{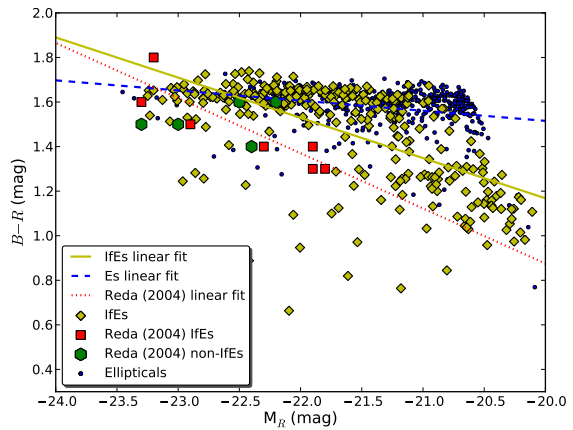}
\caption{Colour ($B-R$) versus absolute $R-$band magnitude for the observed and
simulated IfEs. The data for the Es control sample have been plotted for
comparison. For IfEs and Es a linear fit is shown. For clarity, only every 10th
Es has been plotted. Note the separate population of faint and blue IfEs.}
	\label{colourMag}
\end{figure}

\begin{figure}
\includegraphics[width=84mm]{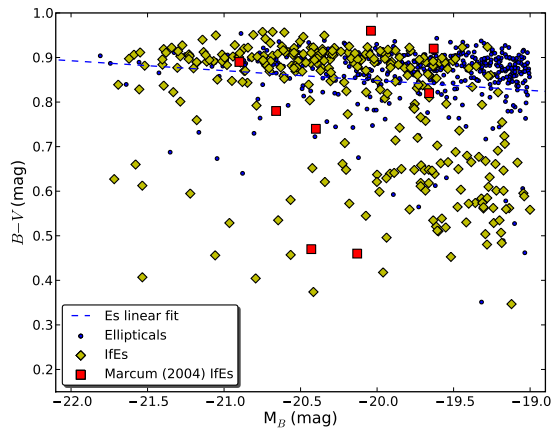}
\caption{Colour ($B-V$) versus absolute $B-$band magnitude for the observed and
simulated IfEs. The data for the Es control sample have been plotted for
comparison. For Es a linear fit is shown. For clarity, only every 10th Es has
been plotted. Note the separate population of faint and blue IfEs.}
	\label{colourMag2}
\end{figure}

The easiest property of IfEs to observe must be the luminosity, and thus the
colour-magnitude relation. This simple yet powerful relation is studied in Figs.
\ref{colourMag} and \ref{colourMag2}, which show the colour-magnitude diagrams
(CMDs) of simulated and observed IfEs. Fig. \ref{colourMag} shows a comparison
between the simulated IfEs and observed field elliptical galaxies presented in
\citet{Reda:2004p502}, while Fig. \ref{colourMag2} shows a comparison to the
sample of IfEs presented in \citet{Marcum:2004p572}. We also plot the Ellipticals
(our control sample) for completeness. Note, however, that in this section we
limit the discussion of the colour-magnitude diagrams to the comparison of
simulated and observed IfEs, while the differences between simulated IfEs and Es
are discussed in Section \ref{ComparisonA}. Colours and associated errors and
standard deviations of observed and simulated IfEs and simulated elliptical
galaxies (Es) are listed in Table \ref{tb:colours}.

Fig. \ref{colourMag} shows that the observed IfEs lie in slightly bluer region in
the CMD than simulated IfEs of the same brightness and that they seem to follow a
tighter correlation than simulated IfEs. The $B - R$ colour scatter of simulated
IfEs is slightly larger compared to observed IfEs: all simulated IfEs show a
$1\sigma$ deviation of $0.23$, however, as the observations of
\citet{Reda:2004p502} are limited to relatively bright ($M_{R} \leq -21.5$)
galaxies, we calculate the colour scatter of bright simulated IfEs using
magnitude cut-off of $M_{R} \leq -21.5$ and find it to be $0.17$. The colour
scatter of the bright simulated IfEs is close to the observed $B - R$ colour
scatter ($0.16$), yet slightly higher. However, one should bear in mind that the
number of observed IfEs is still small.

\begin{table}
\caption{Colours of simulated and observed IfEs and simulated non-isolated
elliptical galaxies (Es).}
\label{tb:colours}
\begin{tabular}{lcclc}
\hline 
Sample & $\Delta mag$ & Mean Colour & Error & $\sigma$\\ 
\hline 
IfEs & $B - R$ & $1.47$ & $0.01$ & $0.23$ \\
IfEs$^{1}$ & $B - R$ & $1.57$ & $0.01$ & $0.17$ \\
IfEs & $B - V$ & $0.79$ & $0.01$ & $0.15$ \\
IfEs$^{2}$ & $B - V$ & $0.82$ & $0.01$ & $0.14$ \\
Es & $B - R$ & $1.58$ & $0.002$ & $0.10$ \\
Es & $B - V$ & $0.86$ & $0.01$ & $0.07$\\
\citet{Reda:2004p502} & $B - R$ & $1.49$ & $0.06$  & $0.16$\\
\citet{Marcum:2004p572} & $B - V$ & $0.76$ & $0.06$ & $0.18$\\
  \hline
\end{tabular}\\
\medskip{Note: The error refers to the standard error of the mean while $\sigma$
is the standard deviation. IfEs$^{1}$ and IfEs$^{2}$ refers to samples where
faint simulated isolated field elliptical galaxies have been removed and the
IfE samples contain only galaxies with $M_{R} \leq -21.5$ and $M_{B} \leq
-19.5$, respectively.}
\end{table}

\citet{Reda:2004p502} found a mean effective colour of $(B - R)_{e} = 1.49 \pm
0.06$ for their isolated elliptical galaxies, which is in good agreement with our
mean value of $1.47 \pm 0.01$ (the errors are the standard error on the mean, see
Table \ref{tb:colours} for details). If we limit our sample of simulated IfEs to
galaxies brighter than $M_{R} \leq -21.5$, as in the sample of
\citet{Reda:2004p502}, the mean $B - R$ colour becomes $1.57 \pm 0.01$. This is
redder than the mean colour in \citet{Reda:2004p502} and shows that our mean
colour is significantly affected by faint IfEs of the blue cloud. When the KS
test is applied to the observed \citep{Reda:2004p502} and all simulated IfE
colours the p-value ($\sim 0.33$) shows that we cannot reject the null hypothesis
at $30$ per cent level. However, if we consider only bright ($M_{R} \leq -21.5$)
simulated IfEs, the p-value is only $\sim 0.01$, implying that a difference
between the observed and simulated IfE $B - R$ colours exists when faint
simulated IfEs are excluded from the comparison.

A straigth line fit to both observed and simulated IfEs differs in the slopes and
intercepts (see Fig. \ref{colourMag}). The linear fit of observed IfEs shows a
steeper slope than the fit of the simulated IfEs. However, as the IfE sample of
\citet{Reda:2004p502} miss galaxies fainter than $M_{R} > -21.5$ it is not
entirely clear what the order of magnitude of this difference might be, and due
to the small number of observed IfEs in the sample of \citet{Reda:2004p502} only
a single new data point at the faint end of the CMD could change the fit
significantly.

The $V-$band (Fig. \ref{colourMag2}) CMD differs from the $R-$band diagram as
correlations seem significantly looser. Here the colour scatter of the observed
IfEs is larger than for the simulated IfEs ($1\sigma$ deviations are $0.18$ and
$0.15$, respectively). The IfE sample of \citet{Marcum:2004p572} extends to
slightly fainter magnitudes than the sample of \citet{Reda:2004p502}, but their
fainter IfEs are surprisingly red ($B - V \sim 0.9$). The simulated IfEs show
slightly redder colours than observed IfEs; the mean $B - V$ colours are $0.79
\pm 0.01$ and $0.76 \pm 0.06$, for simulated and observed IfEs, respectively. The
colours of simulated IfEs agree with observed colours within their standard
errors of the mean, and the KS test approves the null hypothesis with high
probability (p-value $\sim 0.63$) when colours of the simulated IfEs are compare
to the colours of the IfE sample of \citet{Marcum:2004p572}. The p-value would be
only $0.04$ if the observed IfEs of \citet{Marcum:2004p572} were compared against
the simulated Es. Thus, the simulated IfEs agree well with the observed IfEs of
\citet{Marcum:2004p572} when $B - V$ colour samples are compared.

Figs. \ref{colourMag} and \ref{colourMag2} show a separate population of
simulated IfEs that belong to the blue cloud and populate the faint end of the
CMD. Quantitatively the colour and the brightness of this population is
following: $B - R \leq 1.4$ or $B - V \leq 0.7$ and $M_{R} > -21.5$ or $M_{B} >
-20.0$. The population of faint and blue IfEs comprise $\sim 26$ per cent ($76$)
of all simulated IfEs, thus, every fourth IfE belong to this population. We note
that none of the IfEs of \citet{Reda:2004p502} or \citet{Marcum:2004p572}
populates this faint and blue part of the CMDs. The sample of
\citet{Marcum:2004p572} shows two very blue IfEs that are part of the global blue
cloud, but are not part of the population of faint and blue IfEs that is found
from the simulation. However, this might be due to the magnitude limit of
\citet{Marcum:2004p572} sample, as they do not have IfEs fainter than $M_{B} >
-19.5$.

\citet{Marcum:2004p572} found preliminary evidence that $50$ per cent of their
sample of isolated early-type galaxies show blue global colours. However, because
of the small sample size they could not conclude that the higher occurrence of
blue E-type galaxies is related to the extremely low densities of the associated
environments. As we are not limited by a small sample, we can confirm whether
IfEs show bluer global colours than non-isolated early-type galaxies or not. The
mean $B - V$ colour of simulated IfEs is $\sim 0.79 \pm 0.01$ in agreement with
the findings of \citet{Marcum:2004p572}. However, this value is significantly
affected by the population of faint IfEs; if we remove the faint IfEs and
consider only IfEs with $M_{B} \leq -19.5$ the mean colour changes to $\sim 0.82
\pm 0.01$ (see Table \ref{tb:colours} and also Fig.~\ref{f:colour} of the next
Section). This is close, yet slightly bluer, than the mean $B - V$ colour of
simulated ellipticals $(0.86 \pm 0.01)$. A similar result is seen when the $B -
R$ colours are studied; bright $(M_{R} \leq -21.5)$ simulated IfEs are almost as
red as all simulated non-isolated ellipticals (Es) $ B - R \sim 1.57 \pm 0.01$
and $1.58 \pm 0.002$, respectively. Thus, our results show that simulated IfEs do
not show global blue colours if faint IfEs are removed, but the blue colours are
due to the separate, faint and blue population of IfEs. Our theoretical findings
predict that $\sim 26$ per cent of IfEs show global blue colours and that these
IfEs belong to the separate population of faint and blue IfEs.

Different trends in the CMDs, visible in Figs. \ref{colourMag} and
\ref{colourMag2}, may be interpreted as environment effects of galaxy formation.
However, it is also possible that the bluest IfEs have had different formation
mechanism and evolutionary path than redder IfEs. Different evolutionary paths
and merging histories could explain different properties observed at redshift
$z=0$, as well as the influence of environment. Both possibilities will be
explored in detail later.

\subsection[]{Dark Matter Halo Masses}\label{R:masses}

We continue comparisons of simulated and observed IfEs by studying their dark
matter haloes. Even it is far from simple task to derive reliable dark matter
halo masses from observations the comparison is highly interesting, as the dark
matter halo properties and galaxy properties are tightly linked.

Isolated field elliptical galaxies in the MS are mainly found residing inside
dark matter haloes that are lighter than $7 \times 10^{12}h^{-1}$M$_{\odot}$. The
dark matter haloes of simulated IfEs are surprisingly light; the median mass is
only $\sim 1.2 \times 10^{12}h^{-1}$M$_{\odot}$. Even the most massive dark
matter halo hosting an IfE is lighter than $2.2 \times 10^{13}h^{-1}$M$_{\odot}$,
which is comparable to a dark matter halo of a small group.
\cite{Memola:2009p642} calculated the total masses of two of their isolated
ellipticals NGC 7052 and NGC 7785 from X-ray observations and quote values $\sim
5 \times 10^{11}$M$_{\odot}$ and $\sim 1.9 \times 10^{12}$M$_{\odot}$,
respectively. These mass values agree well with our findings of dark matter halo
mass. \cite{Norberg:2008p652} did not find any isolated systems residing in
haloes outside the range $\sim 5 \times 10^{11}$ to $10^{13}h^{-1}$M$_{\odot}$ in
their simulations, therefore, we can conclude that IfEs reside in lighter than
$\sim 2 \times 10^{13}h^{-1}$M$_{\odot}$ dark matter haloes.

The dark matter halo mass also sets constrains to the possible formation
mechanisms; IfEs residing in light haloes cannot be merger remnants of groups or
clusters. Only those IfEs that have a large and massive dark matter halo could
have formed via merger of a group or multiple larger galaxies, but even then the
group should have been poor with only few member galaxies. Evidently, this is
true for fossil groups which are in many ways similar objects to IfEs. The
dynamical masses of fossil groups range from $\sim 10^{13}$ to
$10^{14}h^{-1}$M$_{\odot}$ \citep[see e.g.][]{DiazGimenez:2008p617}.

\subsection[]{Ages}\label{R:ages}

The last property of IfEs we compare is their age. The median mass weighted age
of our model IfEs is $8.84$ Gyr (see Table \ref{tb:properties}). We also note
that the number density of young (mass weighted age less than $5$ Gyr) IfEs is
extremely low (Fig. \ref{f:masswAge}), giving a lower limit for the age of an
IfE.

\citet{Reda:2005p561} found a mean age of their IfEs to be $4.6 \pm 1.4$ Gyr,
while \citet{Proctor:2005p558} quote an age estimate of $\sim 4$ Gyr. These age
estimates are around the young end of our values. However, these estimates should
not be compared directly to our values as the definitions are different: mass-
vs. luminosity-weighted age. Moreover, \cite{Collobert:2006p580} found a broad
range of stellar ages for their IfEs; ranging from $\sim 2$ to $15$ Gyr, in good
agreement with our estimates, as the mass weighted age of model IfEs ranges from
$5$ to $12$ Gyrs. The big scatter suggests that the formation and evolution of
IfEs is not concentrated at a fixed epoch. This can also explain the large
scatter we see in some properties of IfEs, as different formation times and
evolutionary paths can lead to significantly different properties at $z = 0$.

\subsection[]{Environments}\label{R:environment}

To get a complete picture of the properties of isolated field elliptical
galaxies, we have to look at their environment. We define a companion as a galaxy
that resides inside either of the two isolation spheres; the small $0.5h^{-1}$
Mpc or the large sphere $1.0h^{-1}$ Mpc. If not specifically mentioned, the
number of companions refers to the total number of galaxies, $N_{COMP}$, inside
the large isolation sphere. Our definition of a companion does not guarantee that
the galaxy belongs to the same dark matter halo as the IfE. Thus, it is possible
that some of the companion galaxies are not gravitationally bound to the IfE.
Truly isolated galaxies, without any companions in simulations, should be treated
with caution. It may be that the mass resolution is not sufficient to form  small
subhaloes that could harbour a dwarf galaxy.

IfEs can have from $0$ to $\sim 30$ companion galaxies. We do not find a single
IfE with more than $30$ dwarf companions, indicating that our IfEs are in
relatively low density environments. The majority of IfEs have $0$ to $20$
companions, while the mean value of the number of companions, $N_{COMP}$, is 10.7
in the MS. For IfEs with the total number of companion galaxies less than $17$,
most of the companions are found to reside inside the smaller ($0.5h^{-1}$ Mpc)
isolation sphere. As the magnitude difference limit inside the small sphere is
$2.2$ we can be sure that our IfEs (excluding the nine subhalo galaxies discussed
earlier) are well isolated from bright nearby galaxies. So, IfEs do not reside
inside cluster-sized dark matter haloes, with virial radii $\sim 1.5 h^{-1}$ Mpc,
as most their companion galaxies are within $0.5h^{-1}$ Mpc distance from the
halo's main galaxy. If IfEs resided in cluster haloes, we should find companion
galaxies also in the larger isolation sphere. We would also expect to find a
larger number of dwarf companions.

\citet{Reda:2004p502} found that only the very faint dwarf galaxies ($M_{R}
\geq -15.5$) appear to be associated with isolated ellipticals. 
On the contrary, we find companion galaxies with a broad range of magnitudes from
the MS. The mean $R-$band magnitude for these companions is $-17$ mag. The
quartile values for companion $R-$band magnitudes are $-18.2$ and $-15.8$ mag showing that
model IfEs can have relatively bright companion galaxies. Thus, the magnitude gap
between companion galaxies and an IfE seems to be larger in observations than in
simulations. This may be due to the way how galaxy luminosities are treated in
the SAM. Thus, this result is probably not without a bias due to the limiting
mass resolution of the MS.

\begin{figure}
\includegraphics[width=84mm]{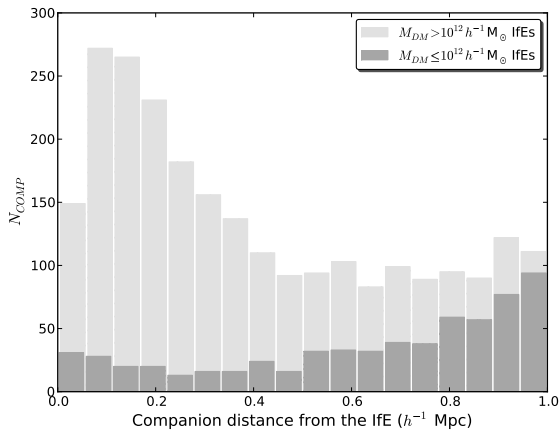}
\caption{Total number of companion galaxies, $N_{COMP}$, inside the large
isolation sphere for light and massive IfEs.}
	\label{f:companionDistance}
\end{figure}

In Fig. \ref{f:companionDistance} we show the distribution of companion galaxy
distances from IfEs. We have divided the IfEs sample into two, to separate the
population of blue, light and faint IfEs; the IfE sample is divided by dark
matter halo mass. Fig. \ref{f:companionDistance} clearly shows that more massive
IfEs have close companions more often than light IfEs. The mean virial radius of
the dark matter haloes of heavy IfEs ($M_{DM} > 10^{12}$M$_{\odot}$) is
$225h^{-1}$ kpc, while it is $115h^{-1}$ kpc for light IfEs ($M_{DM} \leq
10^{12}$M$_{\odot}$). A significant number of companions of heavy IfEs are inside
the mean virial radius and belong to the same dark matter halo as the IfE.
However, for blue, light and faint IfEs the trend is opposite; we find most of
the companions more than six times the mean virial distance away from the IfE.
The median companion distances are $0.33$ and $0.73h^{-1}$ Mpc for heavy and
light IfEs, respectively.

Similar results are also found if the IfEs are divided by their colour. The blue,
$B-R \leq 1.4$, IfEs have most of their companion galaxies more than five times
away than the mean virial radius, while the red IfEs have the majority of their
companions within the mean virial radius. These results show that the most
isolated field elliptical galaxies are blue and relatively faint.

\section[]{Comparison of Regular and Field Ellipticals}\label{ResultsA}

Here we compare the properties of the simulated isolated field elliptical
galaxies to the control sample of all elliptical galaxies in the simulation, to
find out the differences between the two populations. For many properties the
median values are relatively close to each other, but clear differences exist in
the shapes of the distributions. Thus, throughout the following sections we use
the Kolmogorov$-$Smirnov two-sample test to assess whether two distributions are
drawn from the same parent population. Results of the KS tests are given as
probabilities (p-values) and presented in Table \ref{tb:propertiesKSresults}. For
completeness, we also plot the histograms of studied properties (see Figs.
\ref{f:magb} $-$ \ref{f:masswAge}) and present the quartile values in Table
\ref{tb:properties}.

\subsection[]{Morphology}\label{Morph}

The distributions of morphologies ($T$ values) of isolated field elliptical
galaxies and Es are very similar, with median values: $-5.62$ and $-5.80$,
respectively. The KS test approves the null hypothesis with high probability
(p-value $\sim 0.39$), thus, it is likely that the differences in morphology
distributions have arisen by change. This is no surprise as our definition of
IfEs and Es requires that the morphology value is smaller than four.

\subsection[]{Galaxy Colour-magnitude Diagrams}\label{ComparisonA}
The galaxy colour-magnitude diagrams show three main features: the red sequence,
the green valley, and the blue cloud. In general, elliptical galaxies populate
the area known as the red sequence. This is certainly true for most elliptical
galaxies of the MS, however, is this true for IfEs as well?

Figs. \ref{colourMag} and \ref{colourMag2} show that the red sequence of IfEs
starts at $M_{R} \sim -21$ (or $M_{B} \sim -19.5$), while it extends to fainter
magnitudes for Es. A trend of redder colours with increasing luminosity is noted
for both simulated IfEs and Es. However, the slope of the trend is steeper and
the scatter is higher for IfEs. Simulated IfEs show a broader distribution in
colours (Figs. \ref{colourMag} and \ref{colourMag2}) than simulated Es. The
$1\sigma$ scatter of $B - R$ colours is $\sim 0.23$ and $\sim 0.10$ for IfEs and
Es, respectively. For $B - V$ colours the $1\sigma$ values are $\sim 0.15$ and
$\sim 0.07$ for IfEs and Es, respectively. Spearman rank order correlation test
shows a very strong (correlation coefficient $cc \sim -0.65$) correlation for
IfEs (in Fig. \ref{colourMag}) and significant ($cc \sim -0.33$) correlation for
Es. The trends in Fig. \ref{colourMag2} are not as strong according to the
Spearman test: $cc \sim -0.60$ and $\sim -0.27$ for IfEs and Es, respectively.
However, all correlations are highly significant, as the probability to have as
large correlation coefficients for uncorrelated data is $< 10^{-20}$ in all
cases.

The colour correlations in Figs. \ref{colourMag} and \ref{colourMag2} suggest
that IfEs follow a different colour-magnitude trend than Es. The KS tests show
large differences (D-values) when the colours of IfEs are compared to Es. The KS
test rejects the null hypothesis in case of IfEs and Es $B - R$ and $B - V$
colours with high probability (p-values $< 10^{-15}$ in both cases). If we
exclude the separate population of faint and blue IfEs, the previous statement
does not change, only the probabilities (p-value now $< 10^{-6}$). Thus, the
CMDs of IfEs and Es differ significantly even when the separate population of
faint and blue IfEs is removed.

\subsection[]{Colour-mass Diagrams}\label{R:cmassA}

\begin{figure}
\includegraphics[width=84mm]{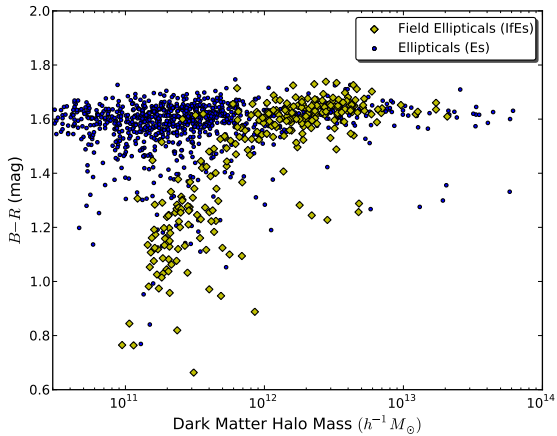}
\caption{Dark matter halo mass versus colour ($B-R$) of the galaxy. For clarity,
only every 5th Es has been plotted.}
	\label{f:colourDmMass}
\end{figure}

Fig. \ref{f:colourDmMass} shows the $B - R$ colour of a galaxy as a function of
the mass of the underlying dark-matter halo. We note from Fig.
\ref{f:colourDmMass} that the Es show a rather constant relation with a few
outliers; in general, Es have the $B-R$ colour $\sim 1.6 \pm 0.15$ independent of
the mass of the dark matter halo they reside in. However, a completely different
trend is seen for IfEs, as they tend to get redder when the dark matter halo mass
grows, while light ($M_{DM} < 10^{12}h^{-1}$ M$_{\odot}$) dark matter haloes host
blue IfEs. The $B-R$ colours of IfEs grow steeply as a function of dark matter
halo mass below $10^{12}h^{-1}$M$_{\odot}$ indicating the influence of dark
matter halo properties to the IfE galaxy they host.

All IfEs are the main galaxies of their dark matter haloes. If instead of the Es
sample we used the MEs for comparison, Fig. \ref{f:colourDmMass} would change.
There would still be a significant number of red, $B-R \sim 1.6$, non-isolated
elliptical galaxies residing in dark matter haloes lighter than $M_{DM} <
10^{12}h^{-1}$ M$_{\odot}$, but their total number would be significantly
smaller. MEs that reside in light ($M_{DM} < 10^{12}h^{-1}$ M$_{\odot}$) dark
matter haloes show a big scatter in colours, indicating that not only the
dark matter halo, but also the environment and the formation history of the halo
effects the galaxy it hosts.

Fig. \ref{f:colourDmMass} further shows that the population of faint and blue
IfEs noted in the colour-magnitude diagrams resides in the lightest dark matter
haloes with $M_{DM} < 10^{12}h^{-1}$ M$_{\odot}$. Below this mass scale we see a
strong correlation between the dark matter halo mass and the evolution of the
galaxy. Thus, simulations predict an unobserved population of IfEs that are blue
and faint and that reside in light dark matter haloes.

\subsection[]{Colour and Luminosity Distributions}\label{R:coloursA}

\begin{figure}
\includegraphics[width=84mm]{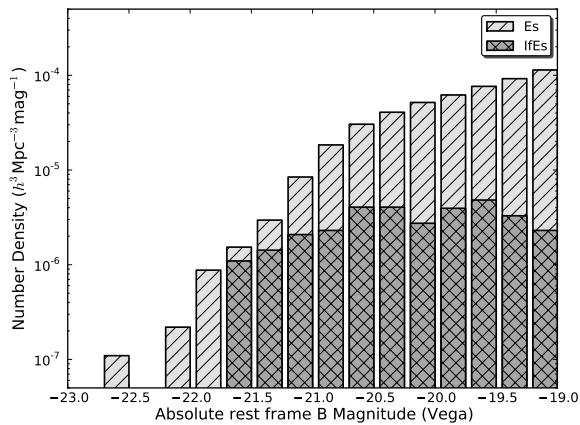}
\caption{Number density of B$-$band rest frame magnitude for isolated field
ellipticals (IfEs) and Es.}
	\label{f:magb}
\end{figure}

Fig. \ref{f:magb} shows the number density of absolute $B-$band rest frame
magnitudes for IfEs and Es. The distributions look very different and the KS test
(Table \ref{tb:propertiesKSresults}) rejects the null hypothesis with high
probability. The differences in the luminosity functions are likely due to the
isolation criteria of IfEs. The most interesting result, seen in the Figure, is
that isolated field elliptical galaxies have an almost constant, $\sim 8 \times
10^{-6}$, number density throughout their $B-$band magnitudes. Surprisingly, when
only brighter ($ -21.7< $ M$_{B} < - 21.0$) elliptical galaxies are considered,
it is almost as probable to find an isolated field elliptical as it is to find a
non-isolated elliptical. It is also noteworthy that we have not identified a
single IfE galaxy brighter than $-21.7$ mag (in $B-$band), while the brightest Es
are almost one magnitude brighter ($\sim -22.5$). This result is natural, since
the brightest elliptical galaxies are normally found to reside in centres of
large clusters. The brightest cluster galaxies (BGCs) have M$_{V} \sim -23.5$ mag
and they have usually experienced many merging events during their evolution
\citep[e.g.][]{DeLucia:2007p414}. Multiple merging events have brought more mass
and gas to the central galaxy and have caused massive star formation and greater
luminosity. It is unlikely that IfEs have followed the same evolutionary path
(see Fig. \ref{f:dmMass} and \ref{f:stMass}).

\begin{figure}
\includegraphics[width=84mm]{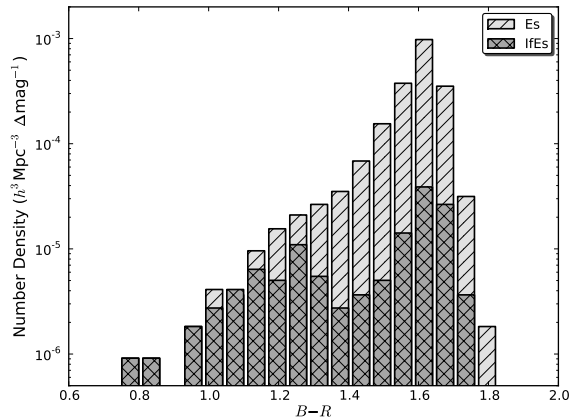}
\caption{Number density of colour distribution ($B-R$ magnitudes) for isolated
field ellipticals (IfEs) and Es.}
	\label{f:colour}
\end{figure}

Since the $B-$band luminosity function of IfEs is almost constant, it leads to
another question, namely if IfEs have a constant number density in colour as
well. Fig. \ref{f:colour} replies to this question and presents the number
density for the B$-$R colour. The distributions for Es and IfEs look very
different and the KS test (Table \ref{tb:propertiesKSresults}) quantifies that
these distributions are not drawn from the same parent population. The isolated
field elliptical galaxies show a bimodal distribution while Es show clearly only
one peak. This result is also seen when the values of first quartiles of both
distributions (Table \ref{tb:properties}) are compared. The value of the first
quartile of our IfEs is almost $0.3$ magnitudes bluer than for the Es in
agreement with \citet{Marcum:2004p572} who found evidence that $50$ per cent of
their sample of isolated early-type galaxies show blue global colours.

The number density of isolated and non-isolated elliptical galaxies is quite
similar for blue, $B-R < 1.3$, galaxies. The bimodality of the IfE
distribution and the blue peak present in Fig. \ref{f:colour} is caused by the
population of blue, faint and light IfEs. The colour distribution of IfEs
suggests that some isolated field ellipticals have either formed later or have
had merger activity at lower redshifts than non-isolated ellipticals. We confirm
whether this is the case in Section \ref{Formation} where formation and merging
times of IfEs are compared to non-isolated ellipticals.

\subsection[]{Dark Matter and Stellar Masses}\label{R:massesA}

We continue our comparisons between IfEs and Es by studying their masses and
composition. Both samples show very little cold gas: $< 1.0 \times 10^{10}h^{-1}$
M$_{\odot}$, with no ongoing star formation (the median star formation rate is
$\sim  0$ M$_{\odot}$yr$^{-1}$). This is typical for elliptical galaxies that are
usually considered as ``red and dead'' at $z = 0$.

\begin{figure}
\includegraphics[width=84mm]{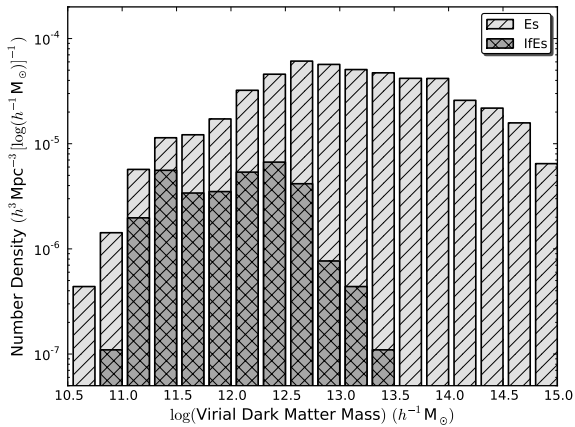}
\caption{Number density of dark matter halo mass for isolated field ellipticals
(IfEs) and Es.}
	\label{f:dmMass}
\end{figure}

When dark matter halo masses are studied in detail, an interesting result is
found. Isolated field elliptical galaxies reside mainly inside dark matter haloes
that are lighter than $7 \times 10^{12}h^{-1}$M$_{\odot}$, while Es are found to
reside more often in more massive haloes. The differences between the dark matter
haloes of Es and IfE galaxies are striking, which is clearly visible in Fig.
\ref{f:dmMass} where we plot the dark matter halo mass distribution as a function
of number density. It is no surprise that the KS test (Table
\ref{tb:propertiesKSresults}) does not approve the null hypothesis, especially as
the distribution of IfEs shows weak bimodality. The great difference in dark
matter halo mass is interesting as the properties of galaxies are tightly related
not only to the environment, but also to the dark matter halo mass within the
galaxy resides in \citep{Croton:2008p600}.



\begin{figure}
\includegraphics[width=84mm]{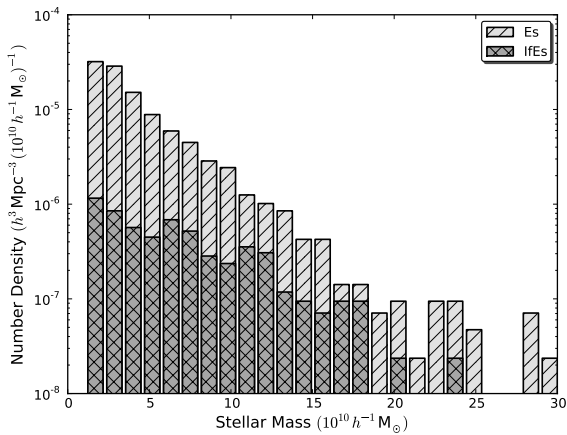}
\caption{Number density of stellar mass for isolated field ellipticals (IfEs) and
Es.}
	\label{f:stMass}
\end{figure}

\begin{figure}
\includegraphics[width=84mm]{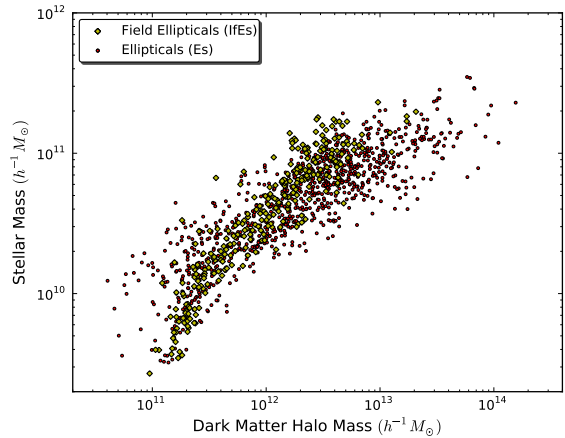}
\caption{A scatter plot of stellar and dark matter halo mass. Yellow diamonds
correspond to the sample of isolated field elliptical galaxies (IfEs), while red
circles correspond to Es. For clarity, only every 2nd galaxy has been plotted.}
\label{f:massComparison}
\end{figure}

While the dark matter haloes of IfEs are, in general, significantly lighter than
haloes of Es, we see a different trend when stellar masses are compared. Even
though the distributions in Fig. \ref{f:stMass} may look somewhat similar, the KS
test (Table \ref{tb:propertiesKSresults}) disproves the null hypothesis with high
probability. The differences in stellar mass distributions become very clear if
we consider the quartile values of both samples. Isolated field elliptical
galaxies have, in general, more stellar mass than Es (Table \ref{tb:properties}).
This is especially intriguing as IfEs reside inside rather light dark matter
haloes (see Fig. \ref{f:dmMass}). To emphasize this difference we plot the dark
matter halo mass versus the stellar mass for both samples in Fig.
\ref{f:massComparison}. IfEs are found to contain more stellar matter in respect
to dark matter than Es. The stellar mass of IfEs grows almost linearly with the
dark matter mass with a mass (dark/stellar) ratio of $\sim 4 \times 10^{-2}$. It
is possible that the difference is related to the formation and evolutionary
paths of IfEs. This possibility is studied in detail in Section \ref{Formation}.

\begin{table}
\caption{Mean values of the properties and results of the Kolmogorov$-$Smirnov
(KS) test: isolated field elliptical (IfEs) vs. control sample (Es) galaxies.}
\label{tb:propertiesKSresults}
\begin{tabular}{lrrc}
  \hline 
  Quantity & IfEs & Es & KS Probability (p-value)\\ 
  \hline 
  Mass$_{DM}$ & $186.29$ & $6751.93$ & $ < 10^{-10}$\\
  Mass$_{ST}$ & $5.46$ & $3.92$ & $ < 10^{-10}$\\
  Mass$_{CG}$ & $0.32$ & $0.10$ & $ < 10^{-10}$\\
  Age & $8.58$ & $9.32$ & $ < 10^{-10}$\\
  Colour & $1.47$ & $1.58$ & $ < 10^{-10}$\\
  M$_{B}$ & $-20.19$ & $-19.76$ & $ < 10^{-10}$\\
  $T$ & $-6.60$ & $-7.06$ & $0.39$\\  
  \hline
\end{tabular}\\
\medskip{Note: Quantity Mass$_{DM}$ refers to virial dark matter mass of the
background halo, Mass$_{ST}$ refers to stellar mass, while Mass$_{CG}$ refers to
the mass in cold gas. All masses are in units of $10^{10}h^{-1}$M$_{\odot}$. Age
is the mass-weighted age of a galaxy in units of $10^{9}$yr. Colour is $B-R$ in
absolute restframe (Vega) magnitudes, and M$_{B}$ is the absolute restframe
(Vega) magnitude in the $B-$band (Buser B3 filter). $T$ is the morphology, and it
is the only quantity for which the null hypothesis of the KS test is approved.}
\end{table}

\subsection[]{Age distributions}\label{R:agesA}

\begin{figure}
\includegraphics[width=84mm]{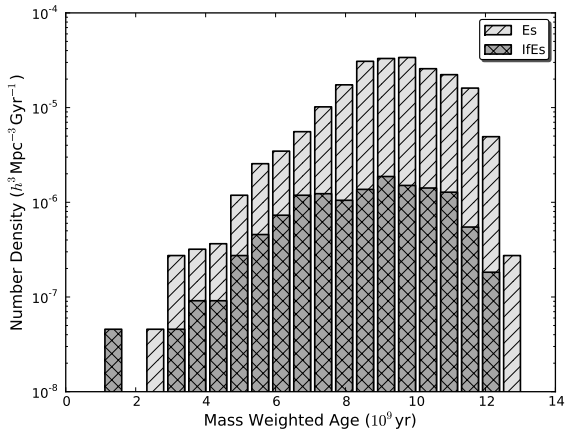}
\caption{Number density of the mass weighted age for isolated field ellipticals
(IfEs) and Es.}
	\label{f:masswAge}
\end{figure}

Distributions of mass weighted age (Fig.~\ref{f:masswAge}) look rather similar,
however, the KS test (Table \ref{tb:propertiesKSresults}) does not approve the
null hypothesis. When quartile values (see Table \ref{tb:properties}) are studied
it is obvious that IfEs have lower mass weighted age than Es, suggesting that
IfEs are younger. Even IfEs seem to be slightly younger than Es in statistical
sense, they cover roughly the same age range, only the extremes are missing.

\begin{table}
\caption{Quartile values of the properties of isolated (IfE) and control sample
(Es) galaxies.}
\label{tb:properties}
\begin{tabular}{lrrrr}
  \hline 
  Sample & Quantity & $1^{st}$ Quartile & Median & $3^{rd}$ Quartile\\ 
  \hline 
  IfEs  & Mass$_{DM}$ & $34.86$ & $120.58$     &    $235.48$\\
  Es & Mass$_{DM}$    &     $311.95$   &     $1069.28$     &   $4998.61$\\
  IfEs & Mass$_{ST}$ &  $1.69$    &       $4.34$      &     $7.94$\\
  Es & Mass$_{ST}$ & $1.97$       &  $2.84$           & $4.73$\\
  IfEs & Mass$_{CG}$ &     $0.10$ &     $0.22$        &   $0.44$\\
  Es & Mass$_{CG}$ & $0.04$       &    $0.05$         &  $0.09$\\
  IfEs & Age     &      $7.21$    &       $8.84$      &    $10.03$\\
  Es & Age       &    $8.44$      &     $9.39$        &  $10.41$\\
  IfEs & Colour  &         $1.28$ &          $1.59$   &        $1.64$\\
  Es & Colour    &       $1.56$   &      $1.60$       &  $1.63$\\
  IfEs & M$_{B}$ &       $-20.70$ &       $-20.18$ &       $-19.64$\\
  Es   & M$_{B}$ &       $-20.14$ &       $-19.65$ &       $-19.28$\\
  IfEs &    $N_{COMP}$ &         $3.00$ &         $7.00$ &        $15.00$\\  
  \hline
\end{tabular}
\medskip{Note: Mass$_{DM}$ is the virial dark matter mass of the background halo,
Mass$_{ST}$ -- the stellar mass, Mass$_{CG}$ -- the mass in cold gas. All masses
are in units of $10^{10}h^{-1}$M$_{\odot}$. Age is the mass-weighted age of a
galaxy in units of $10^{9}$yr. Colour is $B-R$ in absolute restframe (Vega)
magnitudes, and M$_{B}$ is the absolute restframe (Vega) magnitude in the
$B-$band (Buser B3 filter). $N_{COMP}$ is the number of companion galaxies within
the large ($1.0h^{-1}$ Mpc) avoidance sphere around the isolated field elliptical
galaxy.}
\end{table}



\section[]{Formation and Evolution of IfEs}\label{Formation}

The analysis of simulated IfEs in the previous sections shows similarities in
colour-magnitude diagrams and comparable ages, colours, and dark matter halo
masses with observational data. In general, we do not find big discrepancies
between simulations and observations for the properties of IfEs we can compare.
The KS tests between the IfEs and Es do not approve the null hypothesis, except
for morphology. Therefore, IfEs selected with the criteria of Section
\ref{Selection} form a distinct class of objects and are significantly different
from regular elliptical galaxies. As a next step in our analysis we use
simulation data to study evolutionary paths and formation mechanisms of IfEs.
Note that in this section, unless otherwise stated, we use the MEs sample as the
control sample, which contains only those non-isolated elliptical galaxies that
are the main galaxies of their dark matter haloes.

\begin{table*}
\begin{minipage}{150mm}
\caption{Statistics of formation and evolutionary times.}
\label{tb:times}
\begin{tabular}{lcccccccccc}
  \hline 
  Sample & Time	& Mean & Median & $1^{st}$ Quartile & $3^{rd}$ Quartile & Mode &
  Min & Max & Stdev\\
  \hline 
  IfEs & $z_{i}$ & $0.644$ & $0.408$ & $0.183$ & $0.687$ & $0.242$ & $0.064$ &
  $6.200$ & $0.836$\\
  MEs & $z_{i}$ & $1.062$ & $0.624$ & $0.242$ & $1.386$ & $0.116$ & $0.020$ &
  $5.724$ & $1.089$\\
  IfEs & $z_{a}$ & $1.079$ & $0.989$ & $0.564$ & $1.504$ & $1.173$ &
  $0.020$ & $5.289$ & $0.710$\\
  MEs & $z_{a}$ & $1.303$ & $1.276$ & $0.755$ & $1.766$ & $1.276$ & $0.041$ &
  $4.520$ & $0.737$\\
  IfEs & $z_{f}$ & $1.520$ & $1.386$ & $0.828$ & $1.912$ & $1.386$ & $0.012$ &
  $5.289$ & $0.831$\\
  MEs & $z_{f}$ & $1.055$ & $0.687$ & $0.457$ & $1.386$ & $0.564$
  & $0.041$ & $6.197$ &  $0.893$ \\
  MEs$^{*}$ & $z_{f}$ & $3.214$ & $3.308$ & $2.831$ & $3.576$ &
  $3.576$ & $1.276$ & $6.197$ & $0.790$\\
  IfEs & $z_{l}$ & $0.310$ & $0.208$ & $0.116$ & $0.408$ & $0.116$ &
  $0.020$ & $2.070$ & $0.325$\\
  MEs & $z_{l}$ & $0.427$ & $0.230$ & $0.144$ & $0.509$ & $0.242$ &
  $0.020$ & $3.866$ & $0.483$\\
  \hline
\end{tabular}\\
\medskip{Note: Quantities $z_{i}$, $z_{a}$, $z_{f}$ and $z_{l}$ are the
identity, assembly, formation and last merging times, respectively.\\
MEs$^{*}$ refers to a sample where only haloes more massive than $5 \times
10^{12}h^{-1}$M$_{\odot}$ have been considered.}
\end{minipage}
\end{table*}

\subsection{Formation and Evolution Times}

\begin{figure}
\includegraphics[width=84mm]{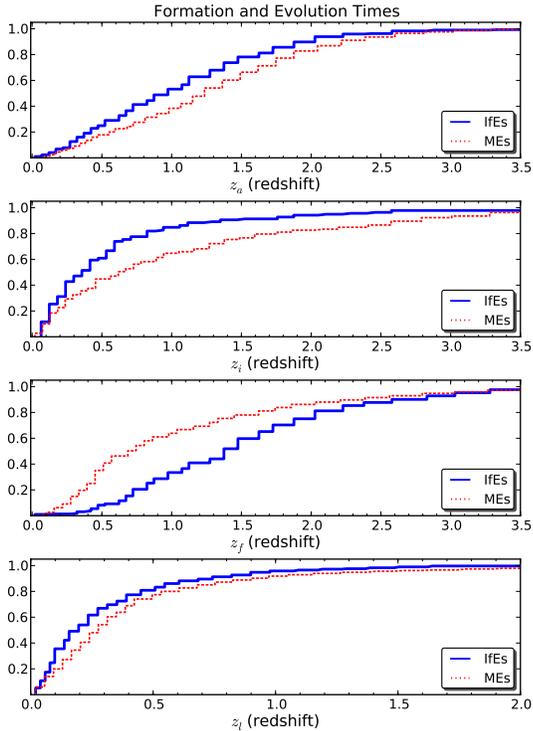}
\caption{Cumulative distributions of assembly ($z_{a}$), identity ($z_{i}$),
formation ($z_{f}$) and last merging ($z_{l}$) times measured in redshifts. The
blue solid lines corresponds to IfEs, while the red dashed lines mark the MEs.}
	\label{f:formationetcTimes}
\end{figure}

\citet{DeLucia:2007p414} defined a set of times related to formation and
evolution of dark matter haloes and galaxies that they reside in. We adopt the
same definitions for convenient and easy comparison, but we add one more quantity
named last merging time, $z_{l}$. Briefly, the different times are defined as
follows:
\begin{itemize}
  \item Assembly time ($z_{a}$) is the redshift when $50$ per cent of the final
  stellar mass is already present in a single galaxy of the merger tree.
  \item Identity time ($z_{i}$) is the redshift when the last major (the two
  progenitors both contain at least $20$ per cent of the stellar mass of the
  descendant galaxy) merger occurred.
  \item Formation time ($z_{f}$) is the redshift when $50$ per cent of the mass
  of the stars in the final galaxy at $z = 0$ have already formed.
  \item Last merging time ($z_{l}$) is the redshift when the last merger
  occurred.
\end{itemize}
We compute all four times, measured in redshifts, related to the formation and
evolution of galaxies, and present numerical results in Table \ref{tb:times}.
Fig. \ref{f:formationetcTimes} shows cumulative distributions of the assembly
($z_{a}$), identity ($z_{i}$), formation ($z_{f}$), and last merging ($z_{l}$)
times. In Table \ref{tb:timeKSresults} we show probabilities that the
distributions of formation and evolution times of IfEs and MEs are drawn from the
same parent distribution. The KS tests indicate great differences in all cases,
implying that the formation and evolution of IfEs is different from MEs.

\begin{table}
\caption{Results of the Kolmogorov$-$Smirnov (KS) test: isolated field
ellipticals (IfEs) vs. the control sample (MEs) galaxies.}
\label{tb:timeKSresults}
\begin{tabular}{ccr}
  \hline 
  Time & D$-$value & KS Probability (p$-$value)\\ 
  \hline 
  	$z_{a}$ & $0.190$ & $5.7 \times 10^{-8}$\\
	$z_{i}$ & $0.239$ & $1.1 \times 10^{-5}$\\
	$z_{f}$ & $0.353$ & $< 10^{-10}$\\
	$z_{l}$ & $0.179$ & $6.5 \times 10^{-8}$\\
  \hline
\end{tabular}\\
\medskip{Note: Quantities $z_{i}$, $z_{a}$, $z_{f}$ and $z_{l}$ are the
identity, assembly, formation and last merging times, respectively.}
\end{table}

Fig. \ref{f:formationetcTimes} (top panel) shows that isolated field elliptical
galaxies have assembled at lower redshifts than galaxies of MEs. Thus, in
general, stars of IfEs form later than stars of MEs and thus IfEs are younger.
This result is natural for hierarchical Cold Dark Matter models. If the
conventional theory, that higher density areas collapse earlier, holds, this
implies that IfEs have formed originally in less dense regions than MEs. Fig.
\ref{f:formationetcTimes} (second panel) also shows that IfEs undergo their major
merging events at significantly lower redshifts than the ellipticals of MEs. The
difference in redshift is significant (see Table \ref{tb:times}) and implies that
a different formation mechanism is behind the evolution of IfEs and MEs.

Fig. \ref{f:formationetcTimes} (third panel) shows that stars that will
eventually form an IfE galaxy are present already at higher redshifts than for
MEs, in agreement with observational findings \cite[e.g.][]{Reda:2005p561}. These
results indicate that IfEs can form stars more efficiently than MEs (see also
Fig. \ref{f:massAssembly}). Note however, that we find the formation time
($z_{f}$) to be highly dependent on the mass of the dark matter halo, as noted in
\citet[][]{DeLucia:2006p376}. If we limit MEs to galaxies with dark matter halo
mass greater than $5 \times 10^{12}h^{-1}$M$_{\odot}$ we find the median
formation time to be at very high redshift ($z \sim 3.3$). This shows that
galaxies in massive dark matter haloes form the bulk of their stars already at
very early cosmic epochs, in agreement with the general hierarchical halo mass
growth scenario.


We define a major merger as an event where the two progenitors both contain at
least $20$ per cent of the stellar mass of the descendant galaxy. Almost half
($\sim 46.3$ per cent) of isolated field elliptical galaxies have experienced a
major merger at some point of their formation history. The percentage of major
merging events for IfEs is higher than for MEs, where only one third ($\sim 33.3$
per cent) of galaxies have experienced a major merging event. If all non-isolated
E-type galaxies, independent of their luminosity or status in their dark matter
haloes, are considered, only about four per cent experience a major merger. The
difference is significant and shows that it is possible to form elliptical
galaxies, isolated or not, without a major merger via disc instabilities.

Let us study the redshifts of the last merging event $z_{l}$ (see Fig.
\ref{f:formationetcTimes} bottom panel and Tables \ref{tb:times} and
\ref{tb:timeKSresults}) and the total number of merging events. IfEs have their
last merging event at lower redshifts than MEs. The median redshift of the last
merging event is $0.21$ and $0.28$ for IfEs and MEs, respectively. This shows
that IfEs have merging activity also at very late stages of their evolution in
agreement with \cite{HernandezToledo:2008p648}. \cite{Hau:1999p629} quoted an age
estimate of $\sim 1.1$ Gyr for isolated elliptical galaxies since the last
merger. This age estimate from observational data seems to disagree with our
median time of the last merging, which is more than twice as high: $\sim 2.5$
Gyr. However, we do find IfEs that have experienced their last merging event only
$\sim 0.3$ Gyr ago. Thus, the discrepancy can arise from the small number of
observed IfEs in \cite{Hau:1999p629}.

It is possible that evolution of IfEs is suppressed by the low density
environment they reside in, thus the major merging events happen later stage in
their evolution if at all. The larger fraction of IfEs having major mergers than
MEs can also be due to the requirement of the magnitude gap not only the region
they formed in. IfEs of denser areas with comparable sized galaxies must clean
their environment effectively, leading to a larger fraction of major merging
events.

While studying the redshifts of the last merging events we noticed that $6$ IfEs
did not experience a single merging event during their evolution. These galaxies
have developed in truly isolated areas; although, maybe, a greater mass
resolution in simulations would reveal one or more minor merging events. Despite
the resolution limitations, the result is in agreement with observations where
some IfEs do not show any signs of merging activity. Typically, simulated IfEs
experience one to $10$ merging events (above the mass limit of
$10^{8}h^{-1}$M$_{\odot}$) during their lifetime, the average being $7.6$ and the
median six mergers. These numbers are significantly higher than for MEs, for
which we find the average and median of $4.6$ and one merging event,
respectively. This suggests that IfEs accrete multiple smaller haloes (and
galaxies) during their formation, cleaning their neighbourhood from dwarf
galaxies effectively. This leads to a question how effectively IfEs accrete mass
as a function of time compared to MEs.

\subsection{Mass Assembly}

\begin{figure}
\includegraphics[width=84mm]{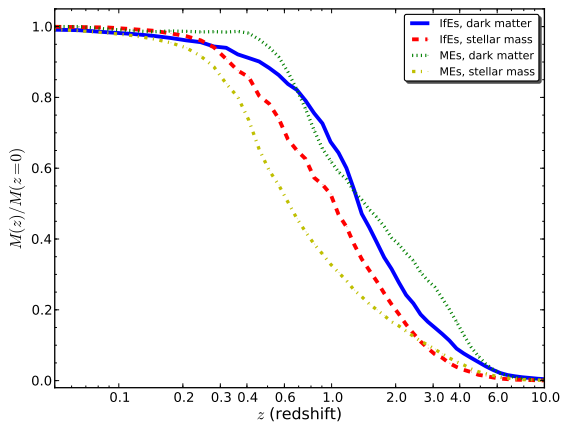}
\caption{Mass assembly of isolated field ellipticals (IfEs) and MEs as a function
of redshift $z$. Lines show the median value at given redshift.}
	\label{f:massAssembly}
\end{figure}

Fig. \ref{f:massAssembly} shows the dark and stellar matter mass assembly as a
function of redshift for both IfEs and MEs. The figure shows that IfEs start to
form around the same epoch as MEs, however, IfEs accumulate stellar matter much
faster. According to our findings IfEs can form stars more efficiently than MEs,
while MEs seem to accrete dark matter slightly faster than IfEs. We confirm that
IfEs form the bulk of their stars at $z > 2$, as suggested in
\cite{Reda:2005p561}. We also note that at redshift one IfEs have formed over
half of their stars (stellar mass) and have gathered as much as $80$ per cent of
their final dark matter. The galaxies of MEs have accreted roughly the same
fraction of dark matter as IfEs at $z \sim 1$, however, they have formed as
little as $30$ per cent of their stars compared to the final stellar matter at $z
= 0$. This difference is significant and shows that IfEs form their stars quickly
and are rather dark-matter poor compared to other elliptical galaxies (see also
Fig. \ref{f:massComparison} for stellar vs. dark matter). It is also noteworthy
that IfEs continue to accrete dark matter till $z = 0$ while MEs have gathered
$\sim 99$ per cent of their final dark matter already at $z \sim 0.5$. All
this points towards a different formation mechanism for these two galaxy classes and
suggests that late merging events can be a significant part of the evolution of
an IfE.

\begin{figure}
\includegraphics[width=84mm]{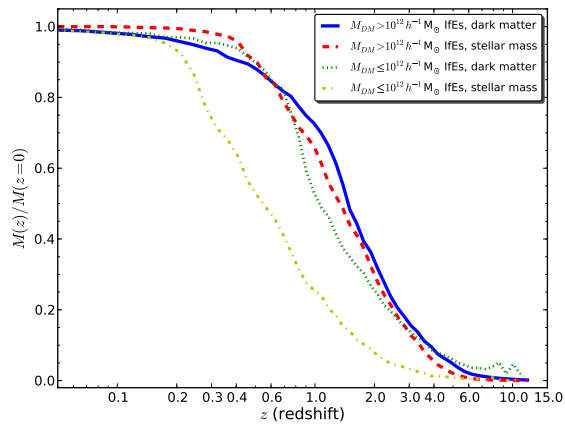}
\caption{Mass assembly of isolated field ellipticals (IfEs) as a function of
redshift $z$ when the sample of IfEs has been divided by dark matter halo mass.
Lines show the median value at a given redshift.}
	\label{f:binnedMassAssembly}
\end{figure}

To further illustrate the mass assembly of IfEs, we plot the sample of blue,
light and faint IfEs separately in Fig. \ref{f:binnedMassAssembly}. We have
divided the IfEs sample into two: light ($M_{DM} \leq 10^{12}h^{-1}$M$_{\odot}$)
and heavy ($M_{DM} > 10^{12}h^{-1}$M$_{\odot}$) IfEs. Fig
\ref{f:binnedMassAssembly} shows that there is a significant difference in mass
accretion of light and heavy IfEs. The stellar mass of heavy IfEs follows closely
the dark matter mass accretion, while light IfEs evolve differently. The heavy
IfEs have created half of their stellar matter already at $z \sim 1.6$, while the
light IfEs have created barely $10$ per cent of their stellar matter, in
agreement with the findings of \cite{Treu:2005p660}. We note that the light IfEs
create half of their stellar matter by $z \sim 0.7$. The heavy IfEs also form
stars extremely efficiently compared to the light IfEs, as the stellar mass
follows dark matter accretion closely. Fig. \ref{f:binnedMassAssembly} also shows
that heavy IfEs host older stellar populations than light IfEs, as $\sim 99$ per
cent of their stellar mass has been accumulated already by $z \sim 0.3$. Thus,
more massive (and luminous) IfEs have old stellar populations, while the lighter
ones have formed a signification fraction of their stellar mass relatively
recently.

\subsection{Merger Histories of IfEs}\label{s:mergerHistories}

So far we have shown that minor and major merging events take place during the
formation of an IfE. Thus, in the following subsections, we try to identify
general formation mechanisms based on different types of merger trees and merging
events. We study merger trees of all IfEs individually and identify three typical
formation scenarios.

We illustrate merger histories in merger plots. The isolated field elliptical
itself lies at the bottom of the plot at $z = 0$, and all its progenitors (more
massive than $10^{8}h^{-1}$M$_{\odot}$) are plotted upwards going back in time.
Galaxies with stellar mass larger than $10^{9}h^{-1}$M$_{\odot}$ are shown as
symbols, and are colour-coded as a function of their rest-frame $B - R$ colour.

\subsubsection[]{Solitude}\label{Form1}

\begin{figure}
\includegraphics[width=84mm]{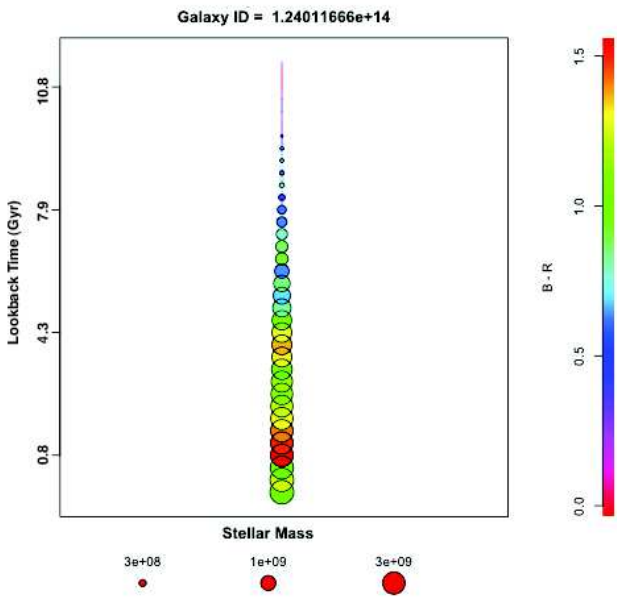}
\caption{Example of a merger tree of an isolated field elliptical galaxy that
has developed alone without any significant merging events. Symbols are
colour-coded as a function of the $B - R$ colour and their area scales with the
stellar mass. Only progenitors more massive than $10^{8}h^{-1}$M$_{\odot}$
are shown.}
	\label{f:alone}
\end{figure}

Fig. \ref{f:alone} shows an example of an isolated field elliptical galaxy that
has developed undisturbed, alone, and has not undergone even a single merging
event. We group IfEs that form in this fashion to a single group that we call
solitude formation class.

Even though the Fig. \ref{f:alone} does not show a single merging event during
the formation history, this may not be the whole story; the plot does not show
mergers smaller than the resolution limit ($\sim 10^{8}h^{-1}$M$_{\odot}$) of the
Millennium Simulation. Therefore, it is possible or even likely that IfEs
belonging to this class have had one or even several minor merging events during
their evolution. Even so, the merging events would have involved very light dark
matter clumps, and it is likely that no significant observational evidence would
exist. In this context small merging events can be interpreted as accretion of
dark matter making solitude a proper formation mechanism.

We identify six IfEs, corresponding to mere two per cent of all IfEs, that belong
to the solitude class. This implies that either IfEs that develop truly alone in
underdense regions are extremely rare or that some of the IfEs have been
misclassified. It is possible that some IfEs with one or more small mergers could
belong to this formation class, especially if the motivation of classification is
based on observability of each class. It is not obvious how massive a merging
event is required to find observational evidence of a merger complicating the
classification.

The six IfEs identified all show unsteady evolution in their colour as a function
of redshift. This is somewhat surprising as one would assume that a passively
evolving galaxy should show a steady colour evolution from blue to red while the
stellar population ages. The bulge formation of solitude class is assumed to
happen via disc instabilities. Solitude IfEs reside inside lighter dark matter
haloes than IfEs of other class, with typical dark matter halo mass of $ \sim 2
\times 10^{11}h^{-1}$M$_{\odot}$ or even less. This makes IfEs that belong to the
solitude class the lightest IfEs in the MS. The formation, when the dark matter
mass grows larger than $10^{8}h^{-1}$M$_{\odot}$, epoch of solitary IfEs is in
the redshift range $3.6 < z < 5.3$. However, we find one solitude IfE that formed
as late as $z \sim 2.5$, making it a very late bloomer, showing that, IfEs can
have formed very recently. Solitude IfEs are in agreement with observations of
e.g. \citet[][]{Aars:2001p564} and \citet[][]{Denicolo:2005p568}, who did not
find any evidence of merging activity while studying their samples of IfEs.


\subsubsection[]{Coupling}\label{Form2}

\begin{figure}
\includegraphics[width=84mm]{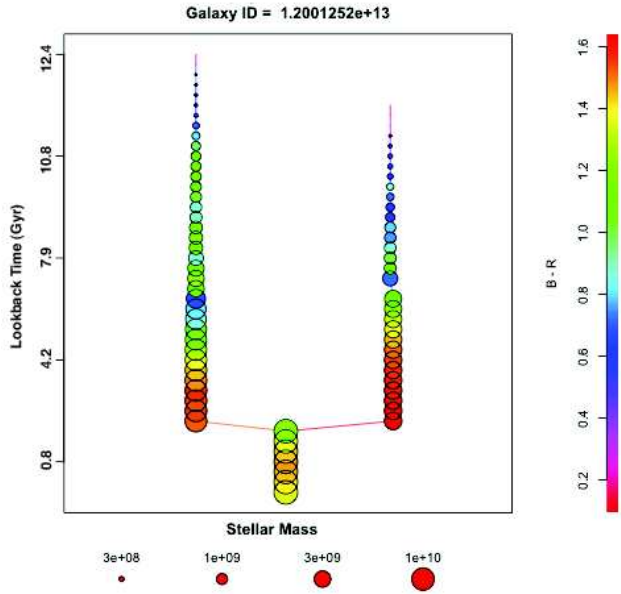}
\caption{Example of a merger tree of an isolated field elliptical galaxy that has
undergone an equal sized merger. Symbols are colour-coded as a function of the
$B - R$ colour and their area scales with the stellar mass. Only progenitors more
massive than $10^{8}h^{-1}$M$_{\odot}$ are shown.}
	\label{f:equal}
\end{figure}

Fig. \ref{f:equal} shows an example of an isolated field elliptical that has
undergone at least one ``equal" size merger during its evolution. IfEs that have
experienced at least one equal sized merger comprise a formation class named
coupling.

Our definition for an equal size merging is the following: the lighter of the
merged galaxies had at least $50$ per cent of the stellar mass of the more
massive one. This is far larger than in our definition of a major merger ($20$
per cent) that was applied when calculating the identity time $z_{i}$. Thus, an
equal sized merger guarantees that the merging event has had a great impact not
only on the morphology, but also on the evolution and other properties of the
IfE.

The motivation for this formation class resides in observations; a major merging
event should be visible in observational data. Thus, an equal sized merging
should leave distinct marks to the descendant galaxy, which should be observable
for at least few Gyrs if not longer. It has been suggested that isolated field
ellipticals have formed in relatively recent mergers of spiral galaxy pairs, i.e.
in merging of two comparable sized galaxies. \cite{Marcum:2004p572},
\cite{Reda:2007p559} and \cite{Kautsch:2008p615} have found several isolated
galaxies that show signs of recent morphological disturbances, while
\cite{Hau:1999p629} \citep[see also][]{Hau:2006p557} have found that $\sim 40$
per cent of isolated galaxies show kinematically distinct cores (KDC). KDCs are
generally believed to be the result of a major or an equal sized merger
\citep{Hernquist:1991p641}. Thus, it is possible that these observations have
already identified galaxies that belong to this formation class. However, it is
also possible that KDCs can form in an early collapse without subsequent mergers
\citep{Harsoula:1998p631}, complicating the identification of the formation
mechanism.

We identify in total $93$ IfEs, corresponding to $\sim 32$ per cent of all IfEs,
that belong to the coupling class of formation scenarios. IfEs that belong to the
coupling formation class show colour evolution that is in agreement with
conventional theory. The time from the last equal sized merging correlates well
with the colour of the galaxy at $z = 0$; galaxies with late merging are bluer
than galaxies that experienced their equal sized merging a long time ago. The
mean redshift of the last equal sized merger is $1.09$ while the median is
$0.62$. IfEs that belong to this formation class start to form usually in the
redshift range $3.6 < z < 5.3$. We also identify a few IfEs that have started to
form as early as $z \sim 8$, making IfEs that belong to the coupling class older,
in a statistical sense, than IfEs that belong to the solitude formation class.

The coupling formation mechanism can explain several observed IfEs. The
colour-magnitude diagrams of \cite{Reda:2004p502} show a slope and scatter that
is in agreement with equal-mass mergers. They find $11$ per cent of their IfEs to
show boxy isophotes that can form when equal sized galaxies merge.
\cite{Marcum:2004p572} argue that all except one of their IfE have luminosities
that would, at most, be consistent with a single equal-mass merger event. Note
that all these observations are best explained by the IfEs of the coupling class.

\subsubsection[]{Cannibalism}\label{Form3}

\begin{figure}
\includegraphics[width=84mm]{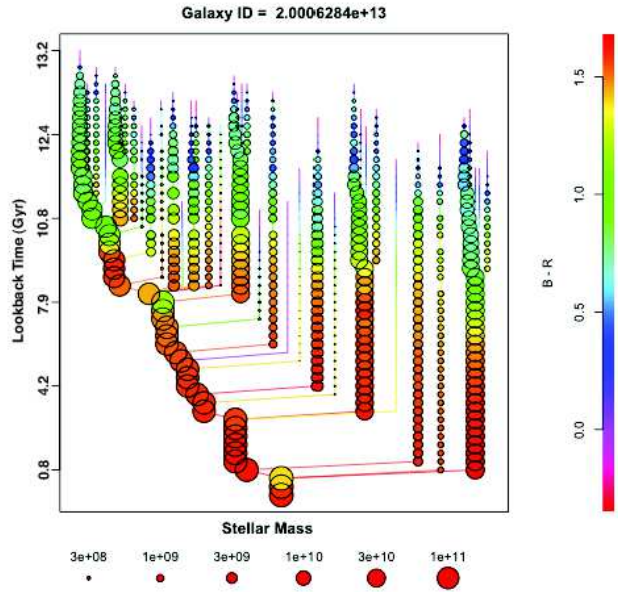}
\caption{Example of a merger tree of an isolated field elliptical galaxy that has
undergone multiple merging events, but not any equal sized ones. Symbols are
colour-coded as a function of the $B - R$ colour and their area scales with the
stellar mass. Only progenitors more massive than $10^{8}h^{-1}$M$_{\odot}$ are
shown.}
	\label{f:cannibal}
\end{figure}

Fig. \ref{f:cannibal} shows a typical isolated field elliptical galaxy that has
developed and accreted dark matter and stellar mass through multiple, small and
larger merging events, but has not experienced any equal sized mergers. All IfEs
that form via multiple mergers form a formation class called cannibalism.

The merger trees of cannibal IfEs show a large number of mergers, with also a
relatively large ones. It is obvious from the merger trees that the colour of the
IfE does not change due to a minor merger. This is due to the fact that many
small merging events are likely to be dry, and therefore, do not induce
significant star formation that would make the global colour of the IfE bluer.
However, the largest merging events can have a significant impact on the IfE's
morphology and colour (see Fig. \ref{f:cannibal} and the merging event around 0.8
Gyr ago). Some merging events, visible in the merger trees of cannibal IfEs, can
be classified as major mergers, and these could be visible in observational data.

Most ($194$) of the IfEs in the MS belong to the cannibal class, corresponding to
$\sim 66$ per cent of all IfEs. The large number of IfEs belonging to this class
shows that merging events are important for the formation and evolution of
isolated field elliptical galaxies. Cannibals often show a large number of
merging events, more than $20$, while the number of major merging events ranges
from one to three. The possibility of major mergers can complicate the
identification of IfEs of this class, especially as we cannot identify any
preferred time for the last major merger. In general, IfEs of the cannibal class
form early, in the redshift range $5 < z < 12$. However, we identify few cannibal
IfEs that form as late as $z \sim 2.4$. The cannibal IfEs that form late could
also be classified as solitudes, especially as these IfEs do not, in general,
show a single major merging event only few minor ones.

This formation class can explain a few observed IfEs. \cite{Reda:2004p502} found
low-luminosity dwarfs close to an isolated galaxy that avoid accretion. Three of
their IfEs show high central space density, but they also found quite strong
morphological disturbances requiring an accretion of a fairly large galaxy. This
is in agreement with cannibal IfEs that have had a major merging event that would
explain the morphological disturbances, while a large number of smaller satellite
dwarfs could have survived due to the large dark matter halo and long dynamical
friction times.

\section[]{Discussion}\label{s:discussion}

\subsection{Population of Blue and Faint IfEs}

Figs. \ref{colourMag} and \ref{colourMag2} show a separate population of blue
and faint IfEs. This population was also confirmed to comprise isolated field
elliptical galaxies that reside in light, $M_{DM} < 10^{12}h^{-1}M_{\odot}$,
dark matter haloes. Moreover, the IfE samples of \citet{Reda:2004p502} and
\citet{Marcum:2004p572} do not reveal a single observed IfE that would belong to
this predicited population of IfEs.

\citet{HernandezToledo:2008p648} studied isolated galaxies and used modern
observations of Sloan Digital Sky Survey (SDSS) Data Release (DR) 6. The base
sample of their galaxies were the Catalogue of Isolated Galaxies (CIG) in the
Northern Hemisphere combiled by \citet{1973AISAO...8....3K}. Although the
isolation criteria in the study of \citet{1973AISAO...8....3K} and in our study
are not the same, it is still interesting to compare if any of the elliptical
galaxies could belong to the population of faint and blue IfEs that our results
predict. \citet{HernandezToledo:2008p648} noted that four early-type galaxies of
the CIG sample showed blue ($g - i < 1.0$) colours. Three of their blue galaxies
were found to be disky while the other one was found to be boxy. From the
morphological inspections they concluded that three of the blue galaxies were
ellipticals while one was S$0$-type. Thus, below we will inspect if any of the
three blue elliptical galaxies noted in \citet{HernandezToledo:2008p648}, could
belong to the faint and blue population of IfEs that simulations predict.

As the magnitudes of \citet{HernandezToledo:2008p648} are in the SDSS system, we
need a method to transform their magnitudes for a comparison. For a crude
comparison we can use galaxy colours of \citet{Fukugita:1995p664} and derive a
transformations:
\begin{equation}\label{eq:r-mag}
M_R \approx M_r - 0.35
\end{equation}
for $R-$band magnitudes for a typical elliptical. Because of our IfEs show bluer
colours than typical ellipticals, this transformation is not very accurate,
however, for our purposes it should be adequate. To transform the SDSS colours,
we use the following equation:
\begin{equation}\label{eq:colour}
B - R \approx (g - i) + 0.44 \quad ,
\end{equation}
which has been derived from the works of \citet{1996AJ....111.1748F} and
\citet{Jester:2005p662}. Again, we note that this may not provide exact colour
transformation, but should provide satisfactory conversion for our crude
comparisons.

The three isolated ellipticals of \citet{HernandezToledo:2008p648} show $g - i$
colours of $0.91$, $0.98$ and $1.02$. With the equation \ref{eq:colour} we can
approximate that these colours correspond to $B - R$ colours of $1.35$, $1.42$,
and $1.46$. Clearly at least one of their galaxy is blue enough ($B - R \leq
1.4$) to belong to the separate population of faint and blue IfEs. Given the
inaccuracy of equation \ref{eq:colour} it is possible that all three of their
galaxies would belong to the blue population of IfEs. The blue galaxies of
\citet{HernandezToledo:2008p648} have absolute $r-$band magnitudes $-20.84$,
$-21.24$, and $-16.73$, respectively. With the help of equation \ref{eq:r-mag} we
can approximate their $R-$band magnitudes to be $-21.19$, $-21.59$, and $-17.08$,
respectively. Thus, at least two of their galaxies are faint enough ($M_{R} >
-21.5$) to be part of the population.

Above magnitudes and colours show that possibly at least one and up to three
isolated field elliptical galaxies that belong to the predicited population of
faint and blue IfEs has already been observed. One of the isolated ellipticals of
\citet{HernandezToledo:2008p648} is significantly fainter than any of our IfEs,
and would not qualify as an IfE in our study. Moreover, as the isolation criteria
of CIG galaxies are different that ours, we cannot conclude without a doubt that
any of the blue and faint population IfEs have been observed. Thus, more
observations of blue and faint isolated elliptical galaxies are required to
confirm our prediction.

\subsection{Observing a Formation Class}

Section \ref{s:mergerHistories} introduced three typical formation classes of
isolated field elliptical galaxies; solitude, coupling, and cannibalism. Our
results show that the majority of IfEs experience numerous merging events during
their evolution and belong to the cannibalism class. A smaller, yet a significant
fraction of IfEs belong to the coupling class, whose members have experienced an
equal sized merging, while only a small fraction of IfEs show insignificant
merging activity and belong to the solitude class. If merging events are so
important for the evolution of IfEs, an obvious question remains: Is it possible
to identify the formation scenario based on observational data? Therefore, in
this section we briefly discuss the properties of IfEs that can be observed and
that at the same time would readily indicate the formation mechanism of the IfE.

IfEs that belong to the coupling class form later than cannibal IfEs and are
found to reside in lighter dark matter haloes than cannibal IfEs: the median
masses are $54$ and $200 \times 10^{10}h^{-1}$M$_{\odot}$, respectively. The
coupling class IfEs can therefore be merger remnants of nearby galaxy pairs
having a modest sized dark matter halo at redshift zero.  Unfortunately, dark
matter haloes cannot be observed directly. IfEs of equal sized mergings can show
a weak X-ray emission; however, it is not clear how well this could be detected
if at all. Moreover, differentiating coupled IfEs from cannibals might be
difficult from X-ray data only.

The mean number of companion galaxies of IfEs that have formed through coupling
is lower than for the cannibal IfEs, but larger than for the solitary IfEs.
Unfortunately, the distribution of stellar mass, $B-$band magnitude and colour of
IfEs of the coupling class is indistinguishable from the cannibal IfEs. The
redshifts of equal sized merging events suggest that to be able to identify
galaxies belonging to the coupling class, observations should concentrate on
intermediate redshifts in the range $0.25 < z < 1.4$. However, to find evidence
of equal sized merging may not be simple; it is not clear how different are the
traces an equal size merger leaves, compared to a major merger. Study of stellar
populations could provide a way to identify IfEs that have experienced an equal
sized merging, but this may not be applicable for high redshifts.

A cannibal IfE can be a merger remnant of a small or compact group. We find
cannibal IfEs to reside in more massive and larger dark matter haloes than other
types of IfEs. However, their dark matter haloes are less massive than haloes of typical
groups, leaving only small or compact groups to consider. Multiple merging events
of cannibal IfEs could also be visible in their stellar populations, especially
if merging events were gas rich, so-called wet mergers. Observational evidence
shows that it is possible to detect IfEs in X-ray observations
\citep[e.g.][]{Mulchaey:1999p563, Sivakoff:2004p628, OSullivan:2004p569,
Memola:2009p642}. Due to their more massive dark matter haloes, X-ray bright IfEs
are the best candidates for the cannibal formation class.


\subsection[]{Comparison to Fossil Groups}

Have fossil groups formed in a similar way as isolated field elliptical galaxies?
Are IfEs an intermediate product while they develop to become fossil groups, or
vice versa? These questions are justified as the selection criteria of IfEs are
very similar to those used for the identification of fossil groups, especially if
only optical data are available. Objects of both classes are required to show a
large magnitude gap between the brightest and the second brightest galaxy in
optical. In general, fossil groups are also required to show extended X-ray
emission while it is not demanded for IfEs. However, several IfEs have shown some
amount of extended X-ray emission even it is not required. Due to similar
selection criteria objects of both classes might have similar evolution and
assembly histories and a similar formation mechanism, being actually the same
class seen only at different phases of evolution.

The majority of fossil groups seem to have experienced the last major merging
event longer than $3$ Gyr ago and they have assembled half of their final mass by
$z \geq 0.8$ \citep{vonBendaBeckmann:2008p598}. Moreover,
\cite{vonBendaBeckmann:2008p598} found from simulations that only $15$ per cent
of fossil groups experience the last major merger less than $2$ Gyr ago, and at
least $50$ per cent had the last major merger longer than $6$ Gyr ago. These
timescales are in modest agreement with our findings. Our results show that IfEs
have experienced their last major merging, on average, $\sim 6$ Gyr ago, while
IfEs have half of their final mass assembled by $z \sim 1$. For fossil groups
this number is $\sim 0.6$ \citep[][]{DiazGimenez:2008p617}.

\cite{DiazGimenez:2008p617} study the evolution of the brightest galaxies of
fossil groups. They identified fossil groups from the Millennium Simulation and
adopted the same galaxy catalogue \citep{DeLucia:2007p414} as in this study,
making it interesting to compare their findings to ours.
When comparing the medians of assembly, formation and identity times we find that
isolated field elliptical galaxies have assembled their stars earlier than the
brightest galaxies of fossil groups; $z_{a} \sim 1.1$ and $\sim 0.6$,
respectively. However, fossil groups form significantly earlier than IfEs; $z_{f}
\sim 3.6$ and $\sim 1.5$, respectively. The median time of the last major merging
of the IfEs is almost twice ($z_{i} \sim 0.6$) the identity time of the brightest
galaxies of fossil groups ($z_{i} \sim 0.3$). These discrepancies in formation
and evolutionary times cast a serious doubt over the idea of similar formation
mechanisms.

\cite{vonBendaBeckmann:2008p598} argue that the primary driver for the large
magnitude gap is the early infall of massive satellites that is related to the
early formation time of fossil groups. We do not find infall of massive
satellites at early time in our sample of simulated isolated field elliptical
galaxies. The early evolution of an IfE includes relatively minor mergers, except
for some IfEs that experience an equal sized merging. However, the time of the
equal sized mergers is usually in a later stage of the development of the IfE,
not early as argued for fossil groups. Our study shows that the primary driver
for the large magnitude gap of IfEs is either a merging of a comparable sized
galaxy pair or effective mass accretion and sweeping up of surrounding galaxies
(coupling and cannibalism, respectively).

\citet{Dariush:2007p265} argue that fossil groups can be identified from dark
matter only simulations if one selects dark matter haloes more massive than $5
\times 10^{13}h^{-1}$M$_{\odot}$. They argue that above that limit, all optical
fossil groups in the Millennium Simulation have enough hot gas to show X-ray
emission, therefore qualifying as X-ray fossils. Our Table \ref{tb:properties}
shows that the median mass of the dark matter haloes of IfEs is only $\sim 1.2
\times 10^{12}h^{-1}$M$_{\odot}$, while even the $3$rd quartile is just $\sim 2.4
\times 10^{12}h^{-1}$M$_{\odot}$. This comparison shows that IfEs reside in
significantly lighter dark matter haloes than the brightest galaxies of fossil
groups. \citet{Dariush:2007p265} argued that fossil groups have formed early and
that more than $\sim 80$ per cent of their mass accumulated as early as $4$ Gyr
ago. Moreover, they suggest that X-ray fossil groups are not a distinct class of
objects but rather that they are extreme examples of groups which collapse early
and experience little recent growth. This formation scenario does not apply for
IfEs, as cannibal IfEs can have merging activity till the redshift $z \sim 0$,
and IfEs start to form later than fossil groups.

Although a large magnitude gap seen in both fossils and IfEs could imply a
similar formation mechanism, the above comparison does not support this idea.
Considering the assembly and formation times it seems unlikely that fossil
groups, and especially their brightest galaxies, could share the same formation
mechanism as isolated field elliptical galaxies. The large differences in dark
matter halo masses of fossil groups and the most massive IfEs makes it improbable
that fossil groups and IfEs are the same class of objects seen at different
phases of their evolution. We therefore conclude that fossil groups and IfEs form
two distinct classes.

\section[]{Summary and Conclusions}\label{summary}

The aim of this paper is twofold: 1) to compare simulated field elliptical
galaxies with observed ones and to make predictions on their properties, and 2)
to define the formation mechanism and history of isolated field elliptical
galaxies. We also discuss if isolated field elliptical galaxies are related to
fossil groups and if they can share a common formation mechanism.

Our results show that it is possible to identify isolated field elliptical
galaxies (IfEs) from cosmological $N-$body simulations with semi-analytical
models of galaxy formation, that have similar properties to observed IfEs. We
show that simulated IfEs are in good agreement with observations when similar
identification criteria are adopted. The colour-magnitude diagram of simulated
IfEs agree with observations, and the average colour of simulated and observed
IfEs are within their error limits when all simulated IfEs are considered.
Unfortunately, observational data sets are still small complicating more detailed
comparison. An agreement in age and mass estimates of IfEs between observations
and simulations is found. However, the age comparisons are less robust due to
different age definitions.

Our results show that isolated field ellipticals are very rare objects; we
find their total number density to be as low as $\sim 8.0 \times 10^{-6}$h$^{3}$
Mpc$^{-3}$. Our result agrees with observational estimates of number of IfEs,
which however, are inaccurate at best. Our IfEs have a small number of companion
galaxies, ranging from only a few dwarf companions to as much as about $20$.
Thus, IfEs are not completely isolated, although they are found to be located in
underdense regions. Our results show that IfEs reside in relatively light dark
matter haloes. However, at the same time, the baryonic to dark matter ratio is
higher in IfEs than in Es generally. The stellar mass of IfEs grows almost
linearly with the dark matter mass with a mass ratio (dark/stellar) of $\sim 4
\times 10^{-2}$, whereas our comparison ellipticals have a lower stellar to dark
matter ratio. Therefore, IfEs are good candidates for galaxies with low dark
matter mass to stellar light ratio. We also find a flat $B-$band luminosity
function for our simulated IfEs.

When studying the basic properties of IfEs we find that IfEs populate different
regions in colour-magnitude diagrams than regular elliptical galaxies, which are
found mostly in the red sequence. This is due to the bimodality of the
distribution of colours, as IfEs populate not only the red sequence, but also the
blue cloud. From simulation data we find a separate population of blue and
faint IfEs. On average IfEs are found to be bluer than our control sample
ellipticals, howevever, this result is biased because of the separate population
of IfEs. The bluest IfEs are found to reside in light dark matter haloes, while
red IfEs are usually found inside more massive ($M_{DM} \geq
10^{12}h^{-1}M_{\odot}$) dark matter haloes. We note that simulations predict a
previously unobserved population of blue, dim and light galaxies that fulfill
observational criteria to be classified as isolated field elliptical galaxies.
These galaxies have only a few companions, which are usually located many times
further away than the virial radius of their dark matter halo. These blue, dim
and light IfEs have formed their stars only lately and have continued to accrete
dark matter mass till redshift zero. This population of IfEs is very interesting
as it has not been detected yet in observations.

Our results also show that IfEs start to form around the same epoch as the
galaxies of the control sample (MEs); however, IfEs seem to accumulate stellar
mass much faster. IfEs form stars more efficiently than MEs, while MEs accrete
dark matter slightly faster than IfEs. We confirm that IfEs form the bulk of
their stars at $z > 2$, as suggested in \cite{Reda:2005p561}. By the redshift one
IfEs have formed over half of their stars (stellar mass) and have gathered as
much as $80$ per cent of their final dark matter. The galaxies of MEs sample have
accreted roughly the same fraction of dark matter as IfEs by $z \sim 1$. However,
they have formed as little as $30$ per cent of their stars compared to the final
stellar matter at $z = 0$. This difference shows that IfEs form their stars
quickly and are rather dark-matter poor compared to other elliptical galaxies.
Moreover, IfEs continue to accrete dark matter till $z = 0$ while MEs have
gathered $\sim 99$ per cent of their final dark matter already by $z \sim 0.5$.
We note that more massive (and luminous) IfEs have older stellar populations,
while the lighter ones have formed a signification fraction of their stellar mass
relatively recently.

While studying the evolution of IfEs we note that almost half ($\sim 46$ per
cent) of IfEs have experienced at least one major merger during their formation
history, while only about four per cent (Es) up to one third (MEs) of control
sample ellipticals experience a major merger. Major merging events happen later
in IfEs evolution than for control sample galaxies, the average of the latest
major merging being $z \sim 0.6$ for IfEs while it is $z \sim 1.1$ for MEs. We
also find IfEs that have not experienced a single merger event above the mass
resolution limit of the MS during their evolution. Therefore, it is possible to
form elliptical galaxies without major mergers.

When inspecting the merger trees of simulated IfEs, we identify three typical
formation scenarios: solitude, coupling, and cannibalism, which can all lead to a
formation of an IfE. The scenarios range from a solitary growth (solitude class)
with quiet mass accretion and star formation to more violent evolution with
multiple mergers (cannibalism class). We also identify a formation scenario where
two comparable sized galaxies merge to form an IfE (coupling class). Our merger
trees show that merging events are important for IfEs that form through
cannibalism or coupling. All three formation classes are in agreement with
observational findings.

Comparison between isolated field elliptical galaxies and fossil groups show that
these two classes are distinct. Galaxies of both classes show some similarities,
but many properties and evolution times are significantly different. IfEs reside
in significantly lighter dark matter haloes and we do not find an infall of
massive galaxies at early times for IfEs as argued for fossil groups
\citep{vonBendaBeckmann:2008p598}. However, we cannot exclude that some fossil
groups could not share the formation mechanism of the most massive IfEs, namely
cannibalism.

\section*{Acknowledgements}

SMN acknowledges the funding by the Finnish Academy of Science and Letters and
the Nordic Optical Telescope Scientific Association (NOTSA). SMN would like to
thank Drs. Elena D'Onghia and Henry Ferguson for enlightening and inspiring
conversations, Dr. Gerard Lemson for invaluable help with the Millennium
Simulation database, and Ms Carolin Villforth for multiple inspiring
conversations. ES thanks the University of Valencia (Vicerrectorado de
Investigaci\'on) for a visiting professorship, the support by the Estonian
Science Foundation grant No. 8005 and by the Estonian Ministry for Education and
Science, grant SF0060067s08. We thank the anonymous referee for detailed reading
of the manuscript and comments that helped us to improve the original manuscript.
The Millennium Simulation databases used in this paper and the web application
providing online access were constructed as part of the activities of the German
Astrophysical Virtual Observatory.

\bsp

\label{lastpage}


\end{document}